\documentclass[fleqn,usenatbib]{mnras}

\usepackage[T1]{fontenc}

\DeclareRobustCommand{\VAN}[3]{#2}
\let\VANthebibliography\thebibliography
\def\thebibliography{\DeclareRobustCommand{\VAN}[3]{##3}\VANthebibliography}

\usepackage{graphicx}	
\usepackage{amsmath}	
\usepackage{amssymb}	
\usepackage{cleveref}
\usepackage{tabularx}
\usepackage[skip=2pt]{caption}
\usepackage{txfonts}

\DeclareMathOperator\erf{erf}

\usepackage{array}
\usepackage{relsize}
\usepackage{acro}
\usepackage{multicol}
\usepackage{multirow}
\usepackage{xfrac}

\DeclareAcronym{illtng}{
  short = TNG ,
  long  = Illustris: The Next Generation ,
  class = abbrev
}

\DeclareAcronym{gqt}{
  short = GQT ,
  long  = Gaussian Quantile Transformation ,
  class = abbrev
}

\DeclareAcronym{cdf}{
  short = CDF ,
  long  = Cumulative Distribution Function ,
  class = abbrev
}

\DeclareAcronym{imf}{
  short = IMF ,
  long  = Initial Mass Function ,
  class = abbrev
}

\DeclareAcronym{sfh}{
  short = SFH ,
  long  = Star Formation History ,
  class = abbrev
}

\DeclareAcronym{zh}{
  short = ZH ,
  long  = Metallicity History ,
  class = abbrev
}

\DeclareAcronym{bgs}{
  short = BGS ,
  long  = Bright Galaxy Survey ,
  class = abbrev
}

\DeclareAcronym{grispy}{
  short = GriSPy ,
  long  = Grid Search In Python ,
  class = abbrev
}

\DeclareAcronym{disperse}{
  short = DisPerSE ,
  long  = Discrete Persistent Structure Extractor ,
  class = abbrev
}

\DeclareAcronym{ghc}{
  short = GHC ,
  long  = Galaxy-Halo Connection ,
  class = abbrev
}

\DeclareAcronym{shmr}{
  short = SHMR ,
  long  = Stellar-Halo Mass Relation ,
  class = abbrev
}

\DeclareAcronym{mzr}{
  short = MZR ,
  long  = Mass-Metallicity Relation ,
  class = abbrev
}

\DeclareAcronym{mhzr}{
  short = HMZR ,
  long  = Halo Mass-Metallicity Relation ,
  class = abbrev
}

\DeclareAcronym{ssp}{
  short = SSP ,
  long  = Simple Stellar Population ,
  class = abbrev
}

\DeclareAcronym{srcc}{
  short = $r_S$ ,
  long  = Spearman's Rank Correlation Coefficient ,
  class = abbrev
}

\DeclareAcronym{mwa}{
  short = MWA ,
  long  = Mass-Weighted Age ,
  class = abbrev
}

\DeclareAcronym{zwa}{
  short = ZWA ,
  long  = Metallicity-Weighted Age ,
  class = abbrev
}

\DeclareAcronym{mpb}{
  short = MPB ,
  long  = Main Progenitor Branch ,
  class = abbrev
}

\DeclareAcronym{rfr}{
  short = RFR ,
  long  = Random Forest Regressor ,
  class = abbrev
}

\DeclareAcronym{ert}{
  short = ERTs ,
  long  = Extremely Randomised Trees ,
  class = abbrev
}

\DeclareAcronym{fof}{
  short = FoF ,
  long  = Friends-of-Friends ,
  class = abbrev
}

\DeclareAcronym{relu}{
  short = ReLU ,
  long  = Rectified Linear Unit ,
  class = abbrev
}

\DeclareAcronym{lrelu}{
  short = L-ReLU ,
  long  = Leaky Rectified Linear Unit ,
  class = abbrev
}

\DeclareAcronym{elu}{
  short = ELU ,
  long  = Exponential Linear Unit ,
  class = abbrev
}

\DeclareAcronym{fsps}{
  short = FSPS ,
  long  = Flexible Stellar Population Synthesis ,
  class = abbrev
}

\DeclareAcronym{desi}{
  short = DESI ,
  long  = Dark Energy Spectroscopic Instrument ,
  class = abbrev
}

\title[Galaxy-Halo Connection With Neural Networks]{Modelling the galaxy-halo connection with semi-recurrent neural networks}

\author[Harry Chittenden \& Rita Tojeiro]{
Harry George Chittenden,$^{1}$\thanks{E-mail: hgc4@st-andrews.ac.uk}
Rita Tojeiro$^{1}$
\\
$^{1}$School of Physics \& Astronomy, University of St Andrews, North Haugh, St Andrews KY16 9SS, Scotland, United Kingdom\\
}

\date{Accepted XXX. Received YYY; in original form ZZZ}

\pubyear{2022}

\begin{document}
\label{firstpage}
\pagerange{\pageref{firstpage}--\pageref{lastpage}}
\maketitle

\begin{abstract}
We present an artificial neural network design in which past and present-day properties of dark matter halos and their local environment are used to predict time-resolved star formation histories and stellar metallicity histories of central and satellite galaxies. Using data from the IllustrisTNG simulations, we train a TensorFlow-based neural network with two inputs: a standard layer with static properties of the dark matter halo, such as halo mass and starting time; and a recurrent layer with variables such as overdensity and halo mass accretion rate, evaluated at multiple time steps from $0 \leq z \lesssim 20$. The model successfully reproduces key features of the galaxy halo connection, such as the stellar-to-halo mass relation, downsizing, and colour bimodality, for both central and satellite galaxies. We identify mass accretion history as crucial in determining the geometry of the star formation history and trends with halo mass such as downsizing, while environmental variables are important indicators of chemical enrichment. We use these outputs to compute optical spectral energy distributions, and find that they are well matched to the equivalent results in IllustrisTNG, recovering observational statistics such as colour bimodality and mass-magnitude diagrams.
\end{abstract}


\section{Introduction}
\label{sec:intro}

The consensual basis for galaxy formation is that galaxies are formed from the contraction of the baryonic gas bound gravitationally to a dark matter halo \citep{Wechsler}. The aspects and future of a galaxy's development are therefore determined by the properties of the halo and its surroundings. While the relationship between certain properties of the halo and galaxy may not necessarily be understood, some degree of correlation between the two is to be expected. Specifically, the rate of change in a host halo's mass, the rate of merger events and similar factors are expected to influence the halo's likelihood of hosting galaxies \citep{Bose}, and the star formation rates, metallicities and other intrinsic properties of the galaxies themselves \citep{Wechsler}. This relationship is commonly referred to as the \ac{ghc}.

The \ac{ghc} holds a number of important scientific questions. In terms of halo and stellar mass growth, the rate at which the halo assembles most of its mass, through accretion and through major mergers, will affect the time at which galaxies form the most stars, become quenched, and cluster together \citep{Croton,Cui,Hani,Montero-Dorta}. In terms of local environment, galaxies in proximity to cosmic filaments are more prone to accreting metal-rich gas \citep{Donnan,PengMaiolino}, and in dense regions of space are tidally quenched by more massive halos, producing noticeably different luminosity and mass functions in their respective environments \citep{Ayromlou, Hellwing, Lu}. This relationship between galaxies and their halos and environments is nonetheless highly complex, and the causal interplay between halos and galaxies, in different cosmic epochs and regimes of halo mass, local density and interaction rates, remains poorly understood.

Several large volume cosmological simulations have, in recent years, synthesised a catalog of halos and galaxies  \citep{Dave, Pillepich2017, Eagle} based on semi-analytic or hydrodynamical modelling of the astrophysical processes which regulate baryonic evolution in fine detail \citep[for reviews of the simulation models see][]{Somerville, Vogelsberger}. However, the complexity of the baryonic models make it computationally impractical to encapsulate the full extent of the \ac{ghc}, and consequently the volumes of such simulations are limited to $ (300 \text{Mpc})^3 $ \citep{IllustrisTNG}, while the larger simulations inevitably limit their mass resolution. The effect of this compromise is problematic in scientific applications, as the rarest of objects, such as high mass clusters, are significantly limited in number and resolution, while their unresolved substructures may contribute significantly to the evolution of causally connected galaxies.

A pure dark matter simulation relies solely on collisionless gravitational and cosmological dynamics, and so can be run on larger volumes without restricting the mass resolution significantly \citep{Vogelsberger, Wechsler}. The halo mass functions, correlation functions and other statistics will appear similar to the equivalent baryonic simulation, given that the dark matter component constitutes the majority of the mass of a galaxy-halo system. A machine learning algorithm which can utilise the properties of a dark matter formation history to emulate their corresponding galaxies can therefore populate a large dark matter simulation with evolving galaxies across all times, in a fraction of the time taken to compute a hydrodynamical simulation of this level of complexity; while the connections of the \ac{ghc} which are learned by this model can offer an explanation of the various galaxy formation mechanisms that take place in the history of the simulation. If applied to a high fidelity N-body simulation, it will result in a vast galaxy dataset with which to test these connections in further detail.

Several studies have used machine learning methods to ascribe galaxy properties to dark matter halos, including physical properties such as stellar and HI mass \citep{Agarwal, JoKim, Lovell}, observational properties such as band magnitudes and galaxy clustering \citep{McGibbonKhochfar, Wadekar2, Xu2}, and their dependence on time and cosmological models \citep{Agarwal, Camels, Xu2}. This study aims to model the complete \ac{sfh} and \ac{zh} of central and satellite galaxies from the historical evolution of their dark matter halos and local dark matter environment by means of a semi-recurrent neural network algorithm. From these properties, we then self-consistently predict observables such as optical spectra and broadband colours. We design two such neural networks: one intended to simulate central galaxies, the other satellite galaxies.

Each network contains two input layers: one is a simple dense layer entailing time-independent halo properties, the second is a simple recurrent layer containing properties defined at multiple time steps. A recurrent layer contains an activation sequence which runs between successive data points, thereby establishing a one-way causal connection between them. For our objective of fashioning an evolutionary history of the dark matter halo and environment, whose temporal properties may play a role in governing the developmental aspects of galaxy formation, this framework proves valuable in enforcing causality and improving the precision of our results.

The motive for developing a semi-recurrent neural network which predicts the star formation rate and stellar mass weighted metallicity over cosmic time is to form our understanding of the dark matter properties which govern galaxy evolution, while determining how accurately a population of galaxies and their historical characteristics can be deduced exclusively from dark matter. Following the development of two neural networks for predicting central and satellite galaxy formation histories, these predictions are used to construct model spectral energy distributions using the \ac{fsps} code \citep{ConroyGunnWhite, ConroyGunn}. We recover observational characteristics, such as the bimodal relationship between stellar mass and colour \citep{Baldry, Cui}.

The neural networks are trained on data from the \ac{illtng} hydrodynamical simulation \citep{Marinacci, Naiman, Nelson2018, IllustrisTNG, Pillepich2017, Springel}, which consists of 100 ``snapshots'' in time, over a redshift domain $0 \leq z \lesssim 20$, with a median time difference of 146 Myr. We access the collaboration's public data repository\footnote{\href{http://www.tng-project.org/data/}{http://www.tng-project.org/data/}}, namely the TNG100-1 and TNG300-1 simulations: the highest resolution simulations with volumes of respective cubic side length 100 Mpc and 300 Mpc. Combining these datasets provides a large, diverse sample of halos of assorted mass and environment, gaining as impartial a training dataset as possible. The \ac{illtng} simulations are based on the Planck-2015 $\Lambda$CDM cosmological model ($\Omega_m = 0.3089$, $\Omega_\Lambda = 0.6911$, $\Omega_b = 0.0486$, $H_0 = 67.74$ km/s/Mpc); assumed throughout this work. To ensure that halo and galaxy formation histories are well resolved at all redshifts, we impose a lower limit on the final stellar mass of $10^9 M_\odot$.

In this paper, we outline the properties, calculation and justification of dark matter quantities in \cref{sec:features}, and baryonic quantities in \cref{sec:baryonic}. Aspects of the neural network design and data preprocessing are discussed in \cref{sec:preprocessing}. We evaluate aspects of the baryonic predictions and their derived observables in \cref{sec:predictions}, and the importance of different features in the model in \cref{sec:relimp}. The implications of the model are reviewed in \cref{sec:discussion} before summarising our findings in \cref{sec:conclusion}.

\begin{table*}
\centering
\begin{tabular}{|c| m{0.21\textwidth} |m{0.09\textwidth}|m{0.1\textwidth} |c|c|c|c|c|} 
 \cline{2-9}
 \multicolumn{1}{c|}{} & \multicolumn{8}{c|}{Network Data} \\ 
 \cline{2-9}
 \multicolumn{1}{c|}{} & \centering Quantity & \centering Notation & \centering Units & Section & Network & GQT & Logarithmic & Shuffle  \\ 
 \cline{2-9} \hline
 \parbox[t]{2mm}{\multirow{15}{*}{\rotatebox[origin=c]{90}{Temporal Features}}}
 & \centering \mbox{Halo Mass} \mbox{Accretion Rate} & \centering $\dot{M_h}$ & \centering $M_\odot/\text{Gyr}$ & \ref{sec:massrates} & Both &  Vector & False & 1  \\ \cline{2-9}
 & \centering \mbox{Subhalo Mass} \mbox{Accretion Rate} & \centering $\dot{m_h}$ & \centering $M_\odot/\text{Gyr}$ & \ref{sec:massrates} & Satellite &  Vector & False & 1a  \\ \cline{2-9}
 & \centering 1Mpc \mbox{Overdensity} & \centering $\delta_1$ & & \ref{sec:overdensity} & Both &  Scalar & False & 2 \\ \cline{2-9}
 & \centering 3Mpc \mbox{Overdensity} & \centering $\delta_3$ & & \ref{sec:overdensity} & Central &  Scalar & False & 2 \\ \cline{2-9}
 & \centering 5Mpc \mbox{Overdensity} & \centering $\delta_5$ & & \ref{sec:overdensity} & Central &  Scalar & False & 2 \\ \cline{2-9}
 & \centering \mbox{Circular Velocity} (proxy) & \centering $\tilde{v}_\text{vir}$ & \centering $\sqrt (M_\odot/\text{Mpc})$ & \ref{sec:vr} & Both &  Vector & False & 3  \\ \cline{2-9}
 & \centering \mbox{Dark Matter} \mbox{Half-Mass Radius} & \centering $R_\frac{1}{2}$ & \centering Mpc & \ref{sec:vr} & Both &  Vector & False & 3  \\ \cline{2-9}
 & \centering 1Mpc \mbox{Radial Skew} & \centering $\mu_3$ & & \ref{sec:skew} & Satellite & Vector & False & 4 \\ \cline{2-9}
 & \centering 3Mpc \mbox{Radial Skew} & \centering $\mu_3$ & & \ref{sec:skew} & Central & Vector & False & 4  \\ \cline{2-9}
 & \centering \mbox{Distance To} \mbox{Closest Subhalo} & \centering $d_{\mu_3}$ & \centering Mpc & \ref{sec:skew} & Both &  Vector & False & 4  \\ \cline{2-9} \hline \hline
 \parbox[t]{2mm}{\multirow{22}{*}{\rotatebox[origin=c]{90}{Non-Temporal Features}}}
 & \centering \mbox{Specific Halo} \mbox{Mass Accretion} Gradient & \centering $\beta$ (c) \newline $\beta_\text{halo}$ (s) & \centering $\log \text{Gyr}^{-2}$ & \ref{sec:beta} & Both &  None & False & 1  \\ \cline{2-9}
 & \centering \mbox{Specific Subhalo} \mbox{Mass Accretion} Gradient & \centering $\beta_\text{sub}$ & \centering $\log \text{Gyr}^{-2}$ & \ref{sec:beta} & Satellite &  None & False & 1a  \\ \cline{2-9}
 & \centering Scaled Infall Time & \centering $a_\text{infall}$ & & \ref{sec:infall} & Satellite &  None & False & 1a, 2, 4  \\ \cline{2-9}
 & \centering \mbox{Scaled Formation} Time & \centering $a_\text{max}$ & & \ref{sec:infall} & Satellite &  None & False & 1a  \\ \cline{2-9}
 & \centering Infall Mass Ratio & \centering $\mu$ & & \ref{sec:infall} & Satellite &  None & True & 1, 1a  \\ \cline{2-9}
 & \centering Infall Velocity & \centering $v_\text{rel}$ & \centering km/s & \ref{sec:infall} & Satellite &  None & True & 2  \\ \cline{2-9}
 & \centering $z=0$ Cosmic Web Distances & \centering $d_\text{CW}$ & \centering kpc & \ref{sec:dfields} & Central &  Scalar & True & 2  \\ \cline{2-9}
 & \centering Starting Time & \centering $t_\text{start}$ & \centering Gyr & \ref{sec:haloage} & Both &  Scalar & False & All  \\ \cline{2-9}
 & \centering \mbox{$z=0$ Halo Mass} & \centering $M_h$ & \centering $M_\odot$ & \ref{sec:mh} & Both &  Scalar & True & 1  \\ \cline{2-9}
 & \centering Maximum \mbox{Absolute Halo} \mbox{Accretion Rate} & \centering $\mid\dot{M_h}\mid$ & \centering $M_\odot/\text{Gyr}$ & \ref{sec:mh} & Both &  Scalar & True & 1  \\ \cline{2-9}
 & \centering \mbox{$z=0$ Subhalo Mass} & \centering $m_h$ & \centering $M_\odot$ & \ref{sec:mh} & Satellite &  Scalar & True & 1a  \\ \cline{2-9}
 & \centering Maximum \mbox{Absolute Subhalo} \mbox{Accretion Rate} & \centering $\mid\dot{m_h}\mid$ & \centering $M_\odot/\text{Gyr}$ & \ref{sec:mh} & Satellite & Scalar & True & 1a  \\ \cline{2-9} \hline \hline
 \parbox[t]{2mm}{\multirow{7}{*}{\rotatebox[origin=c]{90}{Targets}}}
 & \centering \mbox{Star Formation} History & \centering $\mathcal{S}$ & \centering $M_\odot/\text{Gyr}$ & \ref{sec:sfh} & Both &  Vector & False & N/A  \\ \cline{2-9}
 & \centering \mbox{Metallicity} \mbox{History} & \centering $\mathcal{Z}$ & \centering $Z_\odot$ & \ref{sec:zh} & Both &  Vector & False & N/A  \\ \cline{2-9}
 & \centering \mbox{$z=0$ Stellar} \mbox{Metallicity} & \centering $Z$ & \centering $Z_\odot$ & \ref{sec:zh} & Both &  Scalar & True & N/A   \\ \cline{2-9}
 & \centering $z=0$ Stellar Mass & \centering $M_s$ & \centering $M_\odot$ & \ref{sec:sfh} & Both &  Scalar & True & N/A   \\ \cline{2-9}
 & \centering \mbox{Mass Weighted Age} & \centering MWA & \centering Gyr & \ref{sec:sfh} & Both &  Scalar & False & N/A  \\
 \hline
\end{tabular}
\caption{A summary of the quantities used in both neural networks, grouped by layer and ordered by their placement in said layer. This entails the units of each quantity, and indicates which networks utilise them and how they are normalised. The section column indicates which section of this paper discusses this quantity. The shuffle group (final column) indicates which variables are simultaneously scrambled when testing for feature importance (see \cref{sec:relimp2}).}
\label{tab:networks}
\end{table*}

\section{Neural Network Features}
\label{sec:features}

In this section, we discuss the implementation of the various features of the neural network, including how they were calculated, normalised, and expected to benefit the predictability of results. A full summary of the quantities used in this network with details of their preprocessing, implementation and testing is given in \cref{tab:networks}.

\subsection{Time-Dependent Variables}
\label{sec:timedependent}

In the \ac{illtng} simulations, each halo is assigned a merger tree, which contains all progenitor subhalos of the target halo at all prior snapshots of the simulation \citep{Jiang, IllustrisTNG}. A "branch" is defined as a particular path taken by any given subhalo to the present halo, and is defined at all snapshots from the time of the subhalo's formation to the target snapshot. For all objects, we utilise the \ac{mpb}, which is defined as the branch describing the history of the subhalo of the highest mass \citep{IllustrisTNG}. All properties, such as stellar and dark mass components, metallicities and angular momenta, whether applicable to the \ac{fof} or SubFind object, are derived from the \ac{mpb}.

Most variables in our model are normalised by Gaussian Quantile Transformation (see \cref{sec:gqt}); however for temporal variables, there exist two normalisation methods, which we term scalar and vector normalisation (see \cref{sec:vecscanorm}). Scalar normalisation consists of unique transformations for each time step, while vector normalisation applies a single transformation to all data regardless of their time. As discussed in the preprocessing section, and the discussion of temporal quantities where relevant, each of these normalisation methods add certain benefits to the treatment of different quantities.

\subsubsection{Halo Mass Accretion History}
\label{sec:massrates}

The dark matter halo's mass accretion rate is defined using the sum of masses of dark matter particles bound to the \ac{fof} group. We convert this to an accretion rate by finite differencing with respect to the time $t_i$ at each snapshot:
\begin{equation}
\dot{M_h}(t_i) = \frac{M_h(t_i) - M_h(t_{i-1})}{t_i - t_{i-1}}
\label{eq:Mhdot}
\end{equation}

The Subfind subhalo's mass formation rate is defined equivalently:
\begin{equation}
\dot{m_h}(t_i) = \frac{m_h(t_i) - m_h(t_{i-1})}{t_i - t_{i-1}}
\label{eq:mhdot}
\end{equation}

Where necessary, \ac{fof} (halo) and Subfind (subhalo) masses will be respectively denoted $M_h$ and $m_h$ to avoid discrepancy. The use of an accretion rate effectively directs the neural network to recognise its integral over any time interval; the full integral of course corresponding to the halo's final mass (see \cref{sec:mh}). For the sake of treatment of accretion rate as a universal parameter to be integrated over intervals of time, accretion rates are vector normalised.

\subsubsection{Overdensity History}
\label{sec:overdensity}

As a measure of the dark matter environment in proximity to the target halo, we compute and apply the mass-weighted density of halos relative to the simulation mean: the halo overdensity. We compute this for each individual snapshot in order to quantify the environmental history of our halo dataset.

For a given object, the local density is defined as the sum of masses of halos within an arbitrary volume centered on the target's centre of mass, divided by said volume \citep{Agarwal, Bose}, where \ac{illtng} subhalos contribute to the calculation if their centres of mass lie within this volume. The target's local dark matter density, and thus the overdensity, is therefore a function of this volume.

We shall denote the overdensities calculated using a radius of $x$ Mpc as $\delta_x$, i.e. $\delta_1$ for 1 Mpc. These overdensities are used as features in the recurrent input. For central subhalos, we compute overdensities $\delta_1$, $\delta_3$ and $\delta_5$, as each of these will capture environmental structures on different scales. For satellite subhalos, we are interested in smaller scale overdensities as a measure of the state of the halo environment. \citet{Agarwal} deem 200kpc to be a useful overdensity radius for constraining zero-redshift baryonic properties, such as stellar mass and neutral hydrogen fraction. However, through investigating the history of multiple kpc-scale overdensities, while their physical values inevitably differ, as a function of time they are geometrically congruous. Smaller overdensities, however, are more susceptible to Poisson noise. Thus, the 1Mpc parsec overdensity ($\delta_1$) is used as the single overdensity measure for satellites.

We utilise the \ac{grispy} \citep{Grispy} package to compute overdensities. \ac{grispy} is a regular grid processing and nearest neighbour searcher, specifically designed to handle periodic boundary conditions, as is the case with the \ac{illtng} simulations. The "bubble neighbours" query returns the set of objects within a specified distance from the reference coordinate; in our case, halos within xMpc of the centre of mass of each targeted halo.

Unlike the halo mass accretion history, the overdensities are scalar normalised, in spite of being defined at the same time steps. The structure of the local environment is expected to vary to such an extent that the differences in overdensities at successive times are not meaningful, and significantly large that a common quantile transformation can fail to distinguish subsets of large and small value. In fact, vector normalisation of the overdensity values have a strong adverse effect on the quality of predictions. Instead we prioritise the instantaneous environment over its vectorised history, as this is immune to temporal variation in cosmic structure and represents the local density field specific to each halo.

\subsubsection{Radial Dark Matter Skew}
\label{sec:skew}

A third temporal parameter is the mass-weighted radial skew of the distribution of dark matter subhalos, going radially outward from the centre of the target subhalo. For a distribution of variables $x_j$ with weights $w_j$, their statistical moments are given:
\begin{align}
&\mu_1 = \frac{\sum_{j=1}^N w_j x_j}{\sum_{j=1}^N w_j} \\ &\mu_2 = \frac{\sum_{j=1}^N w_j (x_j - \mu_1)^2}{\sum_{j=1}^N w_j} \\ &\mu_n = \frac{\sum_{j=1}^N w_j \left( \frac{x_j - \mu_1}{\sqrt{\mu_2}} \right)^n}{\sum_{j=1}^N w_j}, \ \ \forall n \geq 3
\label{eq:moments}
\end{align}

where the skew is the third moment by definition, and shall henceforth be denoted $\mu_3$. In our implementation, $w_j$ are replaced by the mass of each subhalo, and $x_j$ the distance to the target subhalo, in Mpc.

Merger events between dark matter halos dramatically enhance star formation in the galaxies they host, by introducing cold gas and triggering large tidal disturbances in both merging galaxies. The scale of this is governed by the relative masses of halos (i.e. the merger ratio) and the gas and star content of the galaxies in question \citep{Wechsler, Bose}. Desiring a temporal handle on significant merger events, a number of parameters have been trialled, such as the snapshot of the merger event and the mass ratio at this time. However, merger events take place on varying timescales and thus cannot be assigned a definitive redshift \citep{Rodriguez-Gomez}, and are prone to errors by nuances in the halo referencing system in Illustris \citep{Poole}. In fact, we find that a set of mean merger ratios occurring at sequential time steps fails to constrain the galaxy evolution significantly.

The skew as a function of time offers a parameterisation of the merger history for each halo, as the most massive subhalos will have the largest influence on the distribution of matter, and in the process of a merger, will skew the distribution more and more positively during infall. For satellites, we choose to evaluate the skew out to a 1Mpc radius, aiming to measure both mergers in the central phase and collisions interior to the \ac{fof} halo in the satellite phase. For centrals, we evaluate the skew of the radial distribution up to 3Mpc. This larger radius is necessary to contain data exterior to the largest of \ac{fof} halos, whose accretion activity has a profound effect on the central galaxy. Satellite galaxies, on the contrary, are more dominated by the mass distribution within the \ac{fof} halo, justifying the use of a smaller scale skew measurement.

\begin{figure*}
\includegraphics[width=\linewidth]{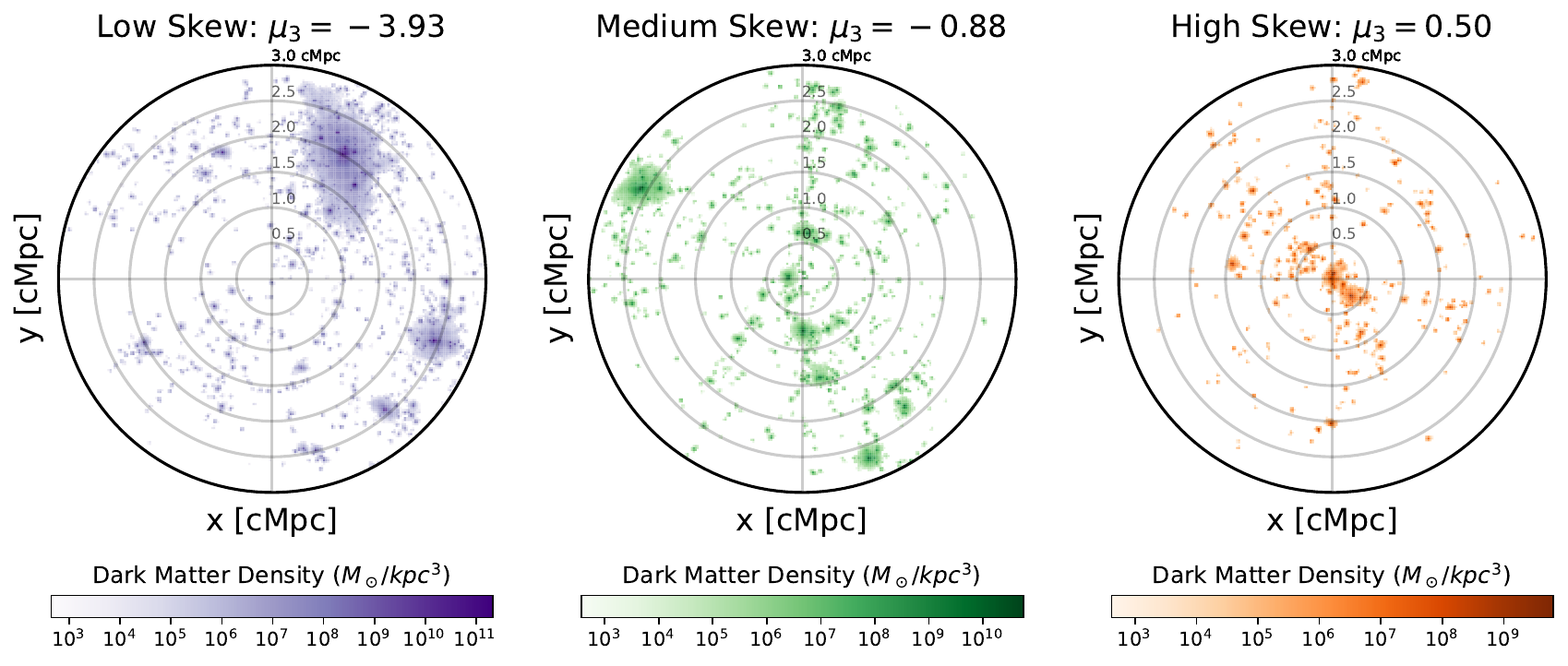}
\vspace*{-20pt}
\caption{Obtained from the z=1 TNG100-1 snapshot, this figure is a visualisation of the x-y plane projection of the dark matter density distribution of subhalos surrounding three central subhalos of $\log M_h^{z=1} / M_\odot = 11.17$. The target subhalos are not shown in these images, just as they play no role in the skew calculation. Dark matter cells are selected for this image provided that they lie within a sphere of radius 3Mpc, centered on the target subhalo's centre of mass, and are not gravitationally bound to the target subhalo. "Low" and "high" skews refer to low and high quantiles of the skew dataset, while "medium" skews are close to the median skew.}
\label{fig:skew}
\end{figure*}

Typical environments for $z=1$ subhalos with low (low quantile), medium (near median) and high (high quantile) skew are exemplified in \cref{fig:skew}, where the target subhalo is omitted for clarity of the exterior mass distributions. An object with insignificant (medium) skew has an unbiased local distribution of matter, such as in the central panel. For a low skew distribution, one or several subhalos, large enough to dominate the local environment, will shift the distribution's centre of mass far from that of the target subhalo. In highly skewed distributions the largest subhalos are instead concentrated near the target subhalo, likely to merge with the target halo or at least invoke a significant tidal disturbance. The skew at a given time is therefore a measure of the local concentration of dark matter, whereas its variation with time describes the nature of flybys and collisions with the subhalo, potentially even its satellites.

As in \cref{sec:overdensity}, the \ac{grispy} package is used to capture the subhalos within a 3Mpc radius of each target subhalo. The IDs and distances from the target to the objects are returned automatically, thus making calculation of the radial mass distribution easy. The skew is simply obtained from this using \cref{eq:moments}, with $n=3$. This skew is computed for all target subhalos, and at all snapshots of \ac{illtng}, to gain a full dataset of the skew histories.

In addition to $\mu_3$, the distance to the closest subhalo is also used for the input of the neural network. Denoted $d_{\mu_3}$ and measured in Mpc, this scales the distribution such that the skew correlates with the true location of the merging halo, and serves as a simple metric for the proximity of a merging halo itself. Both of these quantities are vector normalised.

\subsubsection{Orbital Velocity \& Half-Mass Radius}
\label{sec:vr}

\citet{Lovell} use \ac{ert} to predict zero-redshift baryonic properties from their dark matter halos, and indicate that the maximum circular velocity of the subhalo's rotation curve $(v_\text{max})$ and the radius enclosing half of the subhalo's dark matter mass $(R_{\sfrac{1}{2}})$ have the greatest predicting power in their dataset when concerning stellar, black hole and gas mass components, metallicity, and instantaneous star formation rate; most likely an indicator of the speed of collapse and thus starbursts and black hole accumulation \citep{JJDavies2020, Lovell}. The authors debate the potential use of historical properties in later work, as a means of improved accuracy of their predicted stellar mass function and various mass relations.

A temporal measure of $v_\text{max}$ is not so simple to implement in our model, due to the prescence of baryons influencing the rotation curve. Through inspecting merger trees in \ac{illtng}, these velocities are larger in the hydrodynamical simulations when compared with the N-body equivalent, for objects in the satellite phase and for centrals at high redshift. We compute a term which is unaffected by the prescence of baryons, and is loosely proportional to the temporal virial velocity of the subhalo, in terms of the dark matter half-mass radius $R_{\sfrac{1}{2}}$, and halo and subhalo mass respectively for central and satellite galaxies:
\begin{equation}
\tilde{v}_\text{vir} (t) = \sqrt{\frac{m_h (t)}{R_{\sfrac{1}{2}} (t)}}
\label{eq:virv}
\end{equation}

where we have ignored constant terms such as the Newtonian gravitational constant. This proxy assumes that \ac{illtng} subhalos possess a common radial density profile, and thus a simple scaling between its total mass and the mass enclosed within a given mass radius. As well as an estimate for virial circular velocity, it is similar to the proxy for NFW concentration used for \ac{illtng} data by \citet{Bose}.

We have implemented $R_{\sfrac{1}{2}}$ and $\tilde{v}_\text{vir}$ as features at all redshifts in our dataset. The required quantities are directly obtained from the TNG data catalog, and the features are vector normalised in our preprocessing stage, thereby gaining a description of the growth of the subhalos' physical sizes and rotation curves. 

\subsection{Time-Independent Variables}
\label{sec:timeindependent}

\subsubsection{Final Halo Mass}
\label{sec:mh}

The diversity of galaxy formation with respect to mass is an important one: the highest mass halos, hosting the highest mass centrals and the greatest abundance of satellites, typically exhibit high mass elliptical galaxies, whose metal content is high and whose star formation has ended some considerable time ago. On the contrary, smaller halos host younger, continually star-forming spiral galaxies, with large amounts of metal-poor gas \citep{Wechsler}. The halo mass is therefore an important quantity determining the properties of star formation and chemical enrichment history of our galaxies.

The final halo mass, while taken from the \ac{illtng} data directly, equates to the integral of the halo rate over the full time of the simulation. As is the case with the accretion rates (see \cref{sec:massrates}), the final subhalo dark matter mass is included in the satellite neural network alongside the halo mass, as an indicator of the present-day baryonic properties governed by differing gravitational regimes. Again, the subhalo mass is not considered for centrals due to their tight correlation from the central subhalo being the dominant mass.

As with all variables in \cref{sec:timedependent}, the zero-redshift halo mass is normalised using the \ac{gqt}. The maximum absolute dark matter accretion rate across the halo's history is also used as an input variable, and is normalised the same way. This also serves as a measure of the magnitude of dark matter accretion.

\subsubsection{Specific Rate Gradient}
\label{sec:beta}

\begin{figure}
\includegraphics[width=\linewidth]{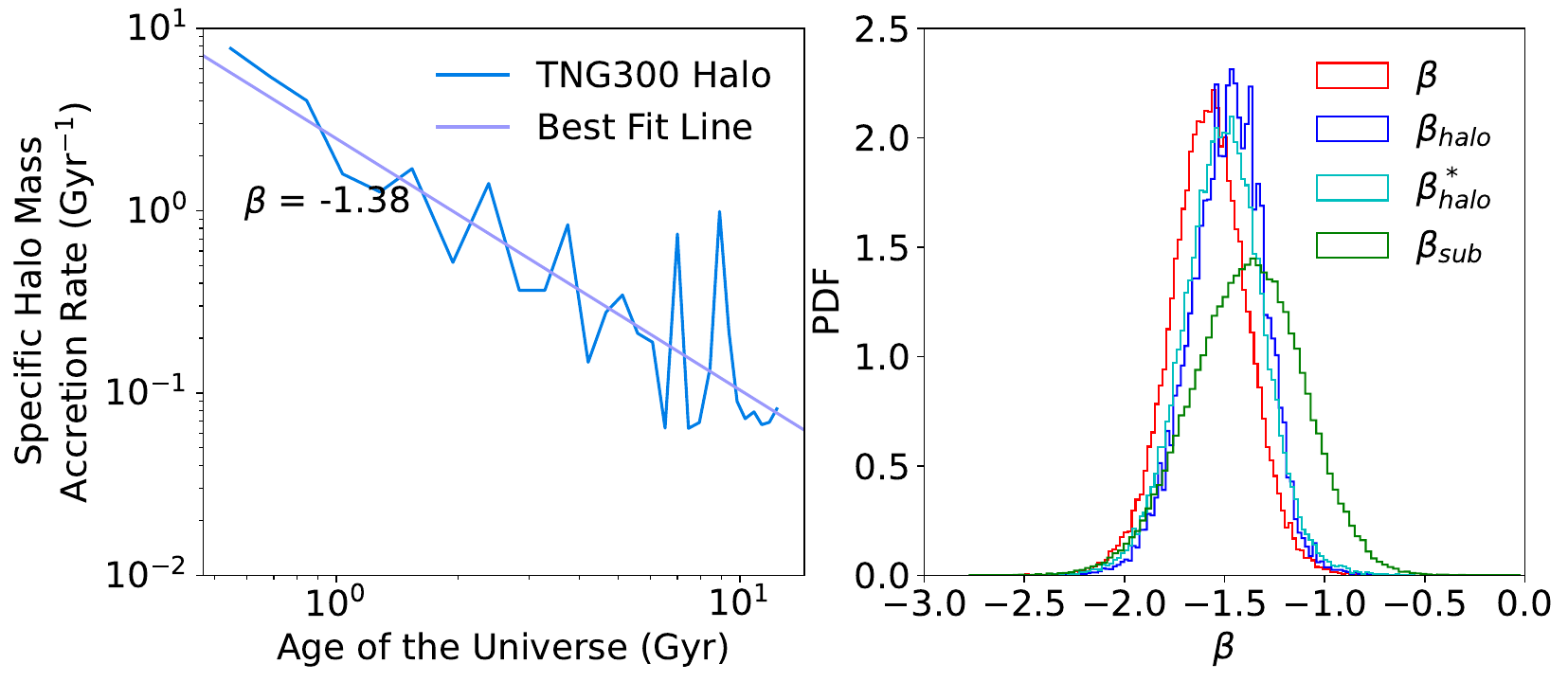}
\vspace*{-20pt}
\caption{(Left) An example of a straight line fit to the specific mass accretion rate of a randomly chosen halo, whose best fit gradient defines $\beta$. (Right) The PDF distributions of our $\beta$ data, including that of the unique subset of the satellite host values, denoted $\beta_\text{halo}^*$. These closely resemble Gaussian distributions and are well fit by the Gaussian function, with best fit parameters given in \cref{tab:beta}. $\beta$ represents the gradients of the halo accretion histories in the central data, the rest are specific to satellite subhalos ($\beta_\text{sub}$) and their host halos ($\beta_\text{halo}$, $\beta_\text{halo}^*$).}
\label{fig:beta}
\end{figure}

A preeminent parameter used by \citet{Montero-Dorta} is the gradient of the specific mass accretion rate, which they denote $\beta$. On the left panel of \cref{fig:beta}, $\beta$ is calculated for one TNG300 halo by fitting a straight line between the logarithms of specific mass accretion rate and cosmic time.

$\beta$ effectively identifies the fastest-forming halos at high redshift, whose galaxies maximise their star formation rate at similar times and go on to form high mass, quenched galaxies. Its inclusion in the neural network advocates a measure of specific accretion, which accounts for the halo's growth in proportion to its current mass, and the time scale of the halo evolving mass fractions. Like the halo mass, this too is an important factor in classifying galaxies of a certain evolutionary regime.

Additionally, we find that this parameter is useful for filtering out outliers, which can negatively impact the network's performance. For the lowest mass objects, in spite of the lower final stellar mass limit of $10^9 M_\odot$, certain accretion histories exhibit a noisy, stochastic appearance as one approaches the mass resolution of the simulation. More importantly, these under-resolved accretion histories are flat, and so their $\beta$ values do not represent a typical specific mass accretion history. We therefore exploit the Gaussian nature of $\beta$ by fitting to its distribution, and discard any samples with over a $5\sigma$ offset from the mean.

As $\beta$ is Gaussian distributed, as can be seen in the right panel on \cref{fig:beta}, its histogram is very similar in geometry to the \ac{gqt} transformed features. Translated according to the best fit parameters, i.e. to a normal distribution of zero mean and unit width, the range of values is also the same as the \ac{gqt} data. Therefore, we simply feed this translated $\beta$, hereafter $\alpha$, into the neural network, with no need for quantile transformation.

For satellites, two distinct halo formation histories serve a role in the model: that of the \ac{fof} halo and that of the satellite subhalo. There are therefore two different $\beta$ values each. Each $\beta$, like the $\beta$ for centrals, is also Gaussian distributed, however for our satellite sample the means are larger and the standard deviations are similar (see \cref{tab:beta}), likely due to the most rapidly forming halos hosting a greater abundance of high-mass satellites.

The $\alpha$ values for each formation history are also included in the satellite neural network, and satellites are selected under the circumstance that neither $\alpha$ has an absolute value above 5. To indicate the halo to which $\alpha$ or $\beta$ refers, these will be denoted $\alpha_\text{halo}/\beta_\text{halo}$ for the main halo, and $\alpha_\text{sub}/\beta_\text{sub}$ for the satellite subhalo.

Note that our satellite data includes multiple subhalos bound to the same halo; there are therefore duplicate values of $\beta_\text{halo}$ in the training set. \Cref{tab:beta} distinguishes the Gaussian fits for the training data $\beta_\text{halo}$ and the unique subset $\beta_\text{halo}^*$, and shows that the true distribution is steeper and more similar to the central $\beta$ distribution, which can be seen in \cref{fig:beta}. Their offset can be attributed to effectively sampling a different mass range; we impose a lower stellar mass cut of $10^9 M_\odot$ to all targets, and for satellites this implies a larger lower boundary of the host mass. 

\begin{table}
\begin{center}
\begin{tabular}{|c|c|c|c|}
\hline
 \multicolumn{4}{|c|}{$\beta$: Best Fit Gaussian Parameters} \\ 
 \hline \hline
 Network & Quantity & Mean & St. Dev. \\ \hline \hline
 Central & $\beta$ & -1.553 & 0.213 \\ \hline \hline
 \multirow{3}{3.4em}{Satellite} & $\beta_\text{halo}$ & -1.489 & 0.192 \\ \cline{2-4}
  & $\beta_\text{halo}^*$ & -1.509 & 0.203 \\ \cline{2-4}
  & $\beta_\text{sub}$ & -1.379 & 0.274 \\ \hline
\end{tabular}
\end{center}
\caption{The best fit mean and standard deviation for all $\beta$ distributions in our data. $\beta_\text{halo}$ has two different fits due to the multiple occurrences of halos which host multiple satellites in our data. $\beta_\text{halo}$ denotes the dataset used in training, while $\beta_\text{halo}^*$ denotes the unique-valued subset of $\beta_\text{halo}$.}
\label{tab:beta}
\end{table}

\subsubsection{Cosmic Web Properties}
\label{sec:dfields}

The \ac{disperse} code \citep{Disperse} is a geometric algorithm which establishes the stationary points of a density field and quantifies its skeletal structure accordingly. We use the cosmic web catalog data built on the \ac{illtng} simulations using \ac{disperse}\footnote{\href{https://github.com/Chris-Duckworth/disperse_TNG}{https://github.com/Chris-Duckworth/disperse\textunderscore TNG/}}. Specifically, we make use of the distances from the target halo to the nearest critical points and dark filaments, denoted collectively as $d_\text{CW}$:

\begin{description}
  \item[$\bullet$ $d_\text{node}$] Distance to the nearest node (maximum) of the density field
  \item[$\bullet$ $d_\text{minima}$] Distance to the nearest void (minimum) of the density field
  \item[$\bullet$ $d_\text{saddle1}$] Distance to the nearest saddle point with one minimised dimension
  \item[$\bullet$ $d_\text{saddle2}$] Distance to the nearest saddle point with two minimised dimensions
  \item[$\bullet$ $d_\text{skel}$] Distance to the midpoint of the nearest filament
\end{description}

This characterisation of the dark matter environment gives a simple description of the level of anisotropy of the large scale environment of the halo, and can be interpreted as characterising the tidal field that surrounds it. The cosmic web has been shown to modulate the accretion of matter onto halos \citep{Borzyszkowski, OHahn}. Observationally, halos of a similar mass will have distinct formation rates according to the surrounding density field \citep{Tinker2018, Tojeiro}, and tend to possess different morphologies and internal dynamics \citep{Hellwing}.

Properties such as halo mass, and scale-independent quantities such as NFW concentration, are also strongly correlated with their tidal environment \citep{Hellwing, Ramakrishnan}, therefore this will not be a unique indicator of the halo formation process. However, the cosmic web does influence the circumgalactic and intergalactic media directly, such as by transfer of metal-rich gas expelled by higher mass galaxies. Thus, they are in principle useful for identifying certain environmental aspects of the galaxy-halo connection.

These cosmic web distances, while useful for modelling the large scale environment influencing accretion of star-forming gas to central halos, is not considered so paramount to objects in the satellite phase. \citet{Simpson} show that the majority of satellite quenching stems from the ram pressure experienced upon infall, or the subsequent tidal effects of the host halo. While they briefly suggest that the cosmic web may quench some satellites, they suggest that this primarily affects low mass satellites which intersect the gas inflow from the filament to the host. Thus, we do not use \ac{disperse} quantities in the satellite neural network.

\subsubsection{Starting Time}
\label{sec:haloage}

Another parameter of the neural network is the time at which the halo first began to form, taken as the earliest snapshot at which the given merger tree is defined.

The structure of the recurrent layer requires identical time steps for all samples, yet halos begin to form at different times. We interpolate the time-dependent properties and return their values at every third snapshot in \ac{illtng}, meaning many will have no data at the earliest times.

While the recurrent layer enforces a causal relationship between time steps, the starting time is nevertheless used as a parameter to establish recently germinated halos directly, and thus, the probable features of their galaxies.

\subsubsection{Satellite Infall}
\label{sec:infall}

The properties of the satellite subhalo and its parent halo at the time of infall being have been shown to be crucial measures of aptitude for star formation \citep{Pasquali,Wetzel}. \citet{Shi} study regimes of satellite evolution by categorising according to a scaled formation time:
\begin{equation}
a_\text{max} \equiv \frac{1 + z_\text{half}}{1 + z_\text{max}}
\label{eq:Shi}
\end{equation}

where $z_\text{max}$ is the redshift of the \ac{illtng} snapshot at which the subhalo's mass is maximised, and $z_\text{half}$ is the redshift of the snapshot at which half of this mass is attained, for the first time. The authors classify satellites as fast-accreting if this value is small, and slow-accreting otherwise; finding that fast-accreting satellites have greater star formation, gas abundance and a systematically distinct \ac{shmr}.

We compute this quantity using the redshifts of the snapshot at which these masses are attained, as \citet{Shi} did in TNG100-1. A second quantity is computed similarly:
\begin{equation}
a_\text{infall} \equiv \frac{1 + z_\text{half}}{1 + z_\text{infall}}
\label{eq:Infall}
\end{equation}

where $z_\text{infall}$ is the redshift at which the subhalo becomes bound to the central halo. $a_\text{infall}$ effectively characterises the stage in the subhalo's growth history at the point of infall, or the timescale of its capture with respect to its growth, whereas $a_\text{max}$, however evidently indicative of satellite galaxy evolution, relates instead to its central-phase growth profile.

A difference in the properties of $a_\text{infall}$ and $a_\text{max}$ is that $a_\text{max}$ is the ratio of two strictly consecutive times in the subhalo's growth, therefore it has a lower bound of 1. For approximately one in 72 of our samples, $a_\text{infall}<1$, indicating that infall occurs before the subhalo reaches half of its peak mass.

Additionally, two physical quantities are evaluated at the time of infall, each of which provide a measure of the properties and dynamics of the halo-subhalo system and how this will affect the future of the tidal environment. One is the absolute velocity of the infalling subhalo relative to its host, calculated simply using the difference in peculiar velocity vectors from each object's merger tree:
\begin{equation}
v_\text{rel}(z_\text{infall}) = \lVert \mathbf{v}_\text{rel}(z_\text{infall}) \rVert = \lVert \mathbf{v}_\text{pec}^\text{sub}(z_\text{infall}) - \mathbf{v}_\text{pec}^\text{halo}(z_\text{infall}) \rVert
\label{eq:relvel}
\end{equation}

In the satellite phase, the velocity of the satellite relative to the halo is related to the rate of the satellite's mass loss from ram pressure, and similarly its orbital velocity serves as a measure of its location in the halo's potential and thus its likelihood of continued star formation \citep{Behroozi, Jiang2}. Including the velocity at the time of infall may prove a useful measure of the satellite's trajectory.

The final infall parameter in this model is the ratio of the subhalo mass to its host halo's mass at infall time:
\begin{equation}
\mu(z_\text{infall}) = \frac{m_\text{sat}(z_\text{infall})}{M_\text{parent}(z_\text{infall})}
\label{eq:muinfall}
\end{equation}

which introduces an important selection criterion which applies only to satellites.

\begin{figure}
\includegraphics[width=\linewidth]{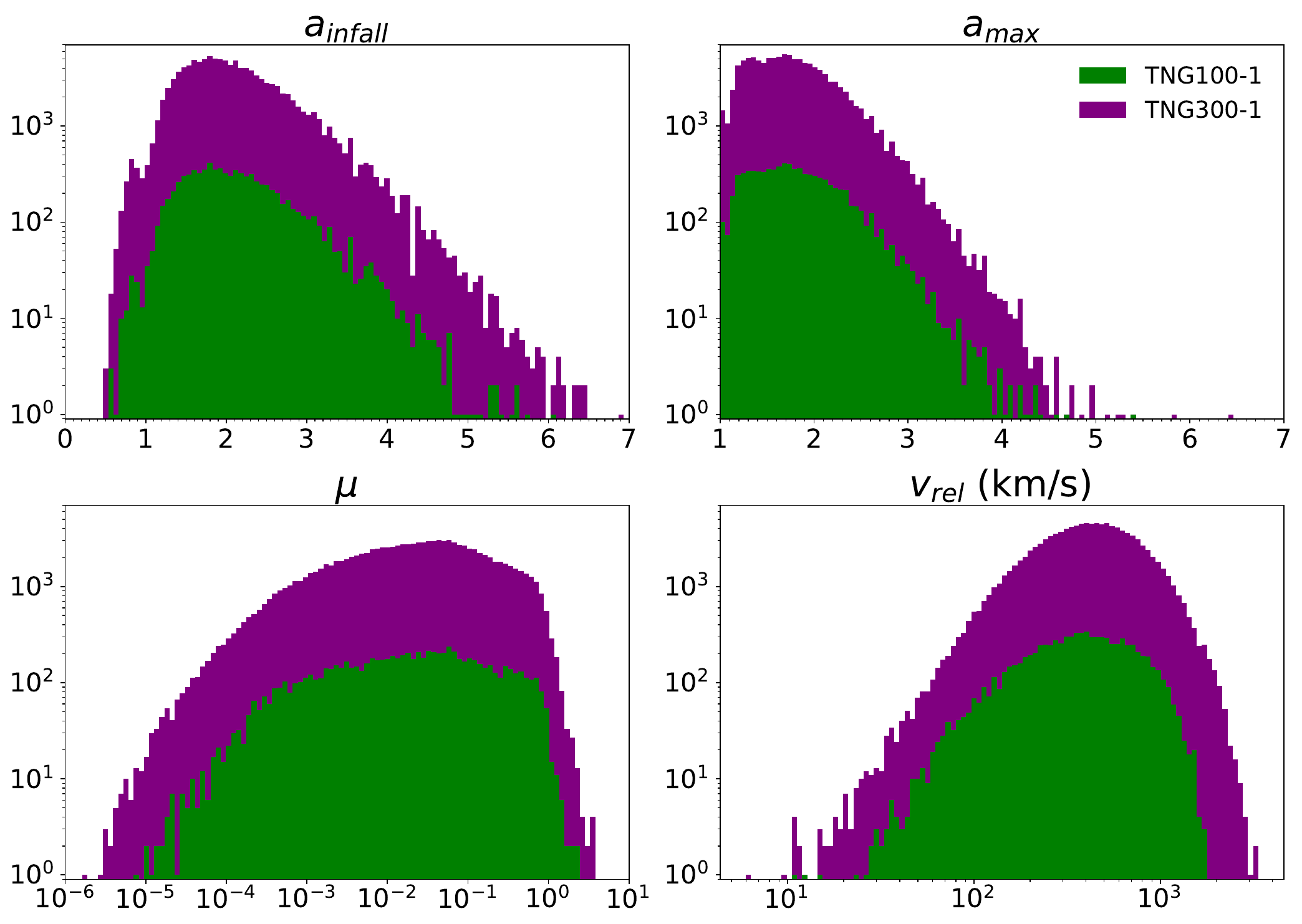}
\vspace*{-20pt}
\caption{Stacked histograms of the four properties of infall satellites, discerned by simulation, used in the neural network for satellites. Clockwise from the top-left, these include the scaled infall time $a_\text{infall}$, the scaled formation time $a_\text{max}$ from Shi et al., the velocity $v_\text{rel}$ of the satellite relative to its host at the infall snapshot, and finally the mass ratio $\mu$ of the two halos.}
\label{fig:infallvar}
\end{figure}

It is assumed that the dark matter subhalo is the dominant mass of the galaxy-halo system. However there exist some low mass subhalos in the \ac{illtng} simulations where the dominant mass is gas or stars, which can be attributed to tidal dwarf galaxies in \ac{illtng} \citep{Haslbauer}. While a satellite may initially be slightly larger than its future central if the latter is rapidly growing, a value of multiple positive orders of magnitude illustrates the collapse of this key assumption. We therefore only include satellite galaxies whose host halo obeys the following criterion: $m_s(z_\text{infall}) / M_h(z_\text{infall})<0.1$, reducing the maximum $\mu$ by two orders of magnitude. The distributions of the infall parameters following this cut are shown in \cref{fig:infallvar}.

\section{Baryonic Quantities}
\label{sec:baryonic}

This section deals with the preprocessing and scientific interpretation of the galaxy properties which our neural network is designed to predict, and their complex role in governing observables.

Like the input variables, these properties are \ac{gqt} normalised. The target dataset consists of the zero-redshift stellar mass and mass-weighted metallicity of the galaxy, and the time-dependent \ac{sfh} and \ac{zh}, both vector normalised and defined for the same time steps as all time-dependent input features.

Defined only on the subhalo scale, the definitions of these historical baryonic quantities are identical for central and satellite galaxies, with the exception of resolution corrections in \cref{sec:zetapsi}, in which the conditional means of these quantities must be evaluated separately for centrals and for satellites. Unlike dark matter properties, we consider only baryonic properties on the subhalo scale.

\subsection{Star Formation History}
\label{sec:sfh}

We define our star formation rate histories from the mass-weighted age distribution of star particles gravitationally bound to each subhalo at $z=0$ - i.e., the stellar mass formed per unit time as a function of cosmic time, $\mathcal{S}(t)$. This definition differs from a stellar mass accretion history, as obtained directly from the merger tree, and analogous to the dark matter accretion rate in \Cref{sec:massrates}. We choose to focus on the first definition given our goal to produce spectral energy distribution for each galaxy. In principle, of course, both can easily be obtained from the simulation.

The total stellar mass ever formed is defined as the integral of the \ac{sfh} $\mathcal{S}$ over time. Due to recycling, this integrated stellar mass is larger than the $z=0$ stellar mass in the merger tree, which we include as a time-independent feature. Unless stated explicitly, this paper refers to the integral of the \ac{sfh} when referring to stellar mass.

We also define a \ac{mwa} for each galaxy as

\begin{equation}
\text{MWA} = \frac{\sum_{j=n}^N \mathcal{M}_n \ t_n^\text{lookback}}{\sum_{n=1}^N \mathcal{M}_n}
\label{eq:mwa}
\end{equation}

where the sum is done over $N$ snapshots, $\mathcal{M}_n$ is the stellar mass formed in snapshot $n$, and $t_n^\text{lookback}$ is the lookback time to snapshot $n$.

We compute the \ac{shmr} for the true and predicted galaxies. This effectively describes the mean stellar mass of a galaxy as a function of its halo mass, while the scatter at fixed halo mass encompasses the variance in \ac{sfh}s associated with differing galaxy growth mechanisms, some of which are highly regulated by halo mass \citep{Behroozi, Gu, Wechsler}. Our model is considered adequately fit to the star formation histories provided that the numerical integrals of their \ac{sfh}s accurately replicate the shape and scatter of the \ac{shmr} at z=0.

\subsection{Stellar Metallicity History}
\label{sec:zh}

The metallicity histories are computed as mass-weighted metallicities of all star particles associated with a subhalo at $z=0$, in the same time bins as the star-formation history.

We also define a mass-weighted metallicity of the full galaxy as:
\begin{equation}
Z_s = \frac{\sum_{n=1}^{N} \mathcal{M}_n \ \mathcal{Z}_n}{\sum_{n=1}^{N} \mathcal{M}_n}
\label{eq:mwz}
\end{equation}

where the sum is done over $N$ snapshots, and $\mathcal{M}_n$ is the mass formed in snapshot $n$. 

\subsection{Resolution Corrections}
\label{sec:zetapsi}

A prominent issue with using the two primary \ac{illtng} simulation datasets is due to their difference in mass and spatial resolution. \citet{Pillepich2017, Pillepich2018} explain that the stochastic star formation model in \ac{illtng} is dependent on the density of gas mass which is identified, therefore simulations with lower resolution underestimate the star formation rate overall. Consequently, important summary statistics such as the \ac{shmr} and \ac{mzr} are underestimated in TNG300-1 in contrast with TNG100-1.

\begin{figure}
\includegraphics[width=\linewidth]{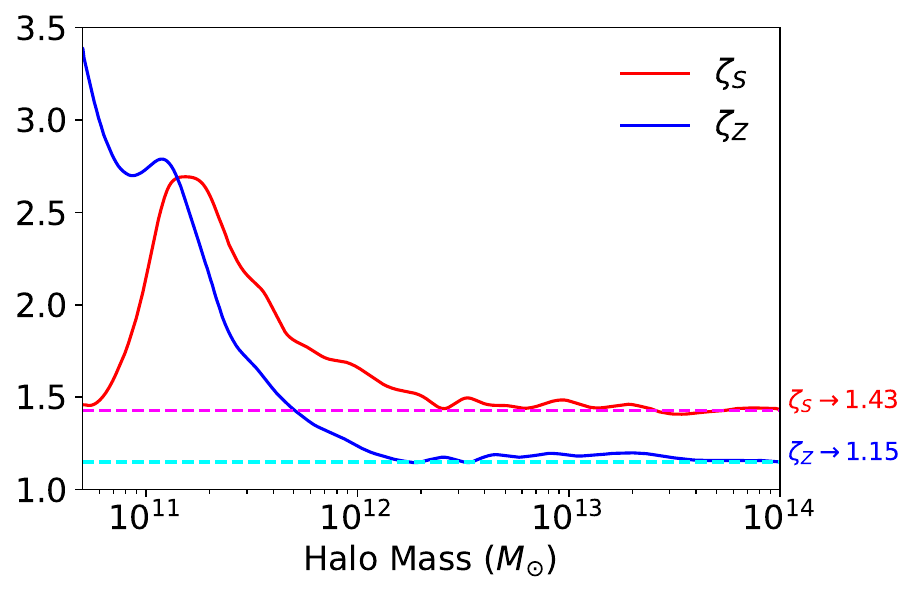}
\vspace*{-20pt}
\caption{The $\zeta$ variables as a function of halo mass at redshift zero. The values of each $\zeta$ function for masses larger than that shown here (i.e. $> 10^{14} M_\odot$) are taken as the average of the function over the $[10^{13}M_\odot, \ 10^{14}M_\odot] $ interval. The lower mass limit of this chart equates to the cutoff of TNG300 halos used in our data.}
\label{fig:zeta}
\end{figure}

The same authors offer an adjustment to TNG300-1 by exploiting the fact that the resolution of TNG100-2 is identical to it. More generally, the $n^\text{th}$ TNG300 simulation is chosen to recover the mean \ac{shmr} of the $n+1^\text{th}$ TNG100 simulation. While they utilise this only to correct the \ac{shmr}, we will do the same to adjust our stellar metallicities. These corrections are based on the halo mass dependent ratio between the mean \ac{shmr}s and mean \ac{mzr}s of TNG100-1 and TNG100-2, and we denote them $\zeta_S$ and $\zeta_Z$, respectively. The shapes of these functions of halo mass at redshift zero are depicted in \cref{fig:zeta}.

At a fixed redshift, the $\zeta$ fractions are defined:
\begin{align}
\zeta_S(M_h \mid z) = \bar{M^{*}}_\text{100-1}(M_h) \ / \ \bar{M^{*}}_\text{100-2}(M_h) \\ \zeta_Z(M_h \mid z) = \bar{Z^{*}}_\text{100-1}(M_h) \ / \ \bar{Z^{*}}_\text{100-2}(M_h)
\label{eq:zeta}
\end{align}

where due to a lack of high mass samples, $\zeta_S$ and $\zeta_Z$ are assigned their mean value in the interval $10^{13} M_\odot \leq M_h \leq 10^{14} M_\odot$ if $M_h \geq 10^{14} M_\odot$.

The $\zeta$ corrections work adequately for adjusting the relevant relations within a single snapshot. However, the stellar age spectra used to compute the formation histories of the network are not defined with single-snapshot data, and so $\zeta$ is not a valid correction for these. Instead we apply a similar correction for the mean histories of objects in narrow bins of final halo mass, over which we apply a cubic spline interpolation, which can be neatly extrapolated outside the halo mass range of TNG100.

\begin{figure}
\includegraphics[width=\linewidth]{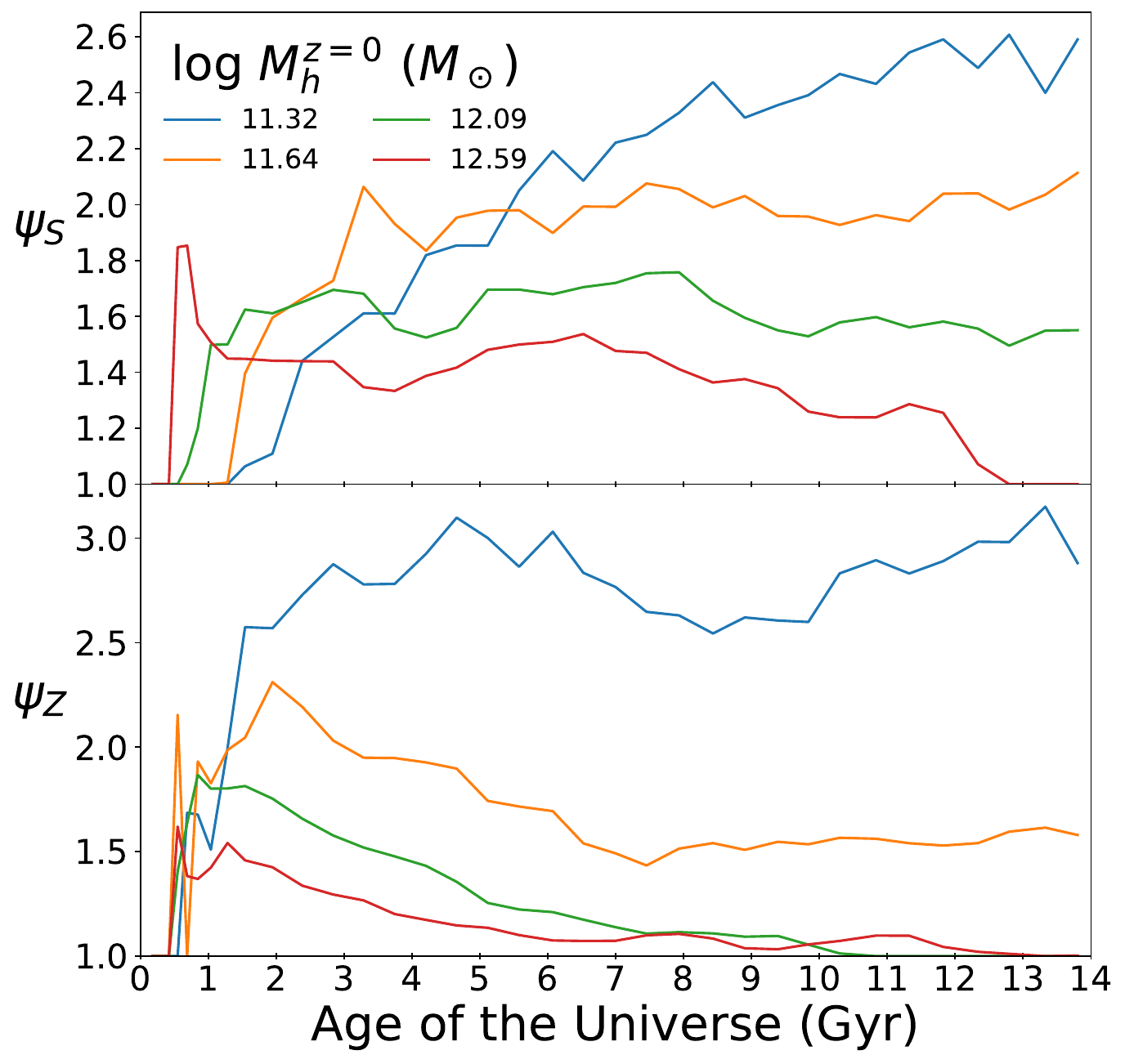}
\vspace*{-20pt}
\caption{The $\psi$ variables as a function of cosmic time, shown in four example bins of halo mass. The reslution correction for the star formation history $\psi_S$ is shown in the upper panel, the metallicity history correction $\psi_Z$ in the lower panel.}
\label{fig:psi}
\end{figure}

This returns a temporal variable, named $\psi$, whose geometry for star formation and metallicity as a function of both time and halo mass is realised in \cref{fig:psi}.

For a fixed halo mass, $\psi$ is given mathematically as follows:
\begin{align}
& \psi_S(z \mid M_h^{z=0}) = \tilde{\mathcal{S}}_\text{100-1}(z) \ / \ \tilde{\mathcal{S}}_\text{100-2}(z) \\ & \psi_Z(z \mid M_h^{z=0}) = \tilde{\mathcal{Z}}_\text{100-1}(z) \ / \ \tilde{\mathcal{Z}}_\text{100-2}(z)
\label{eq:psi}
\end{align}

To reiterate, the baryonic properties required for this study include the final stellar mass $M_s$ and metallicity $Z_s$, whose TNG300 values are multiplied by $\zeta$: a dimensionless function of halo mass at fixed redshift; and the star formation history $\mathcal{S}$ and metallicity history $\mathcal{Z}$, whose TNG300 values are multiplied by $\psi$: a dimensionless function of redshift at fixed final halo mass. $\zeta$ and $\psi$ are computed independently for central and satellite galaxies, due to significant differences in their star formation histories.

\begin{figure}
\includegraphics[width=\linewidth]{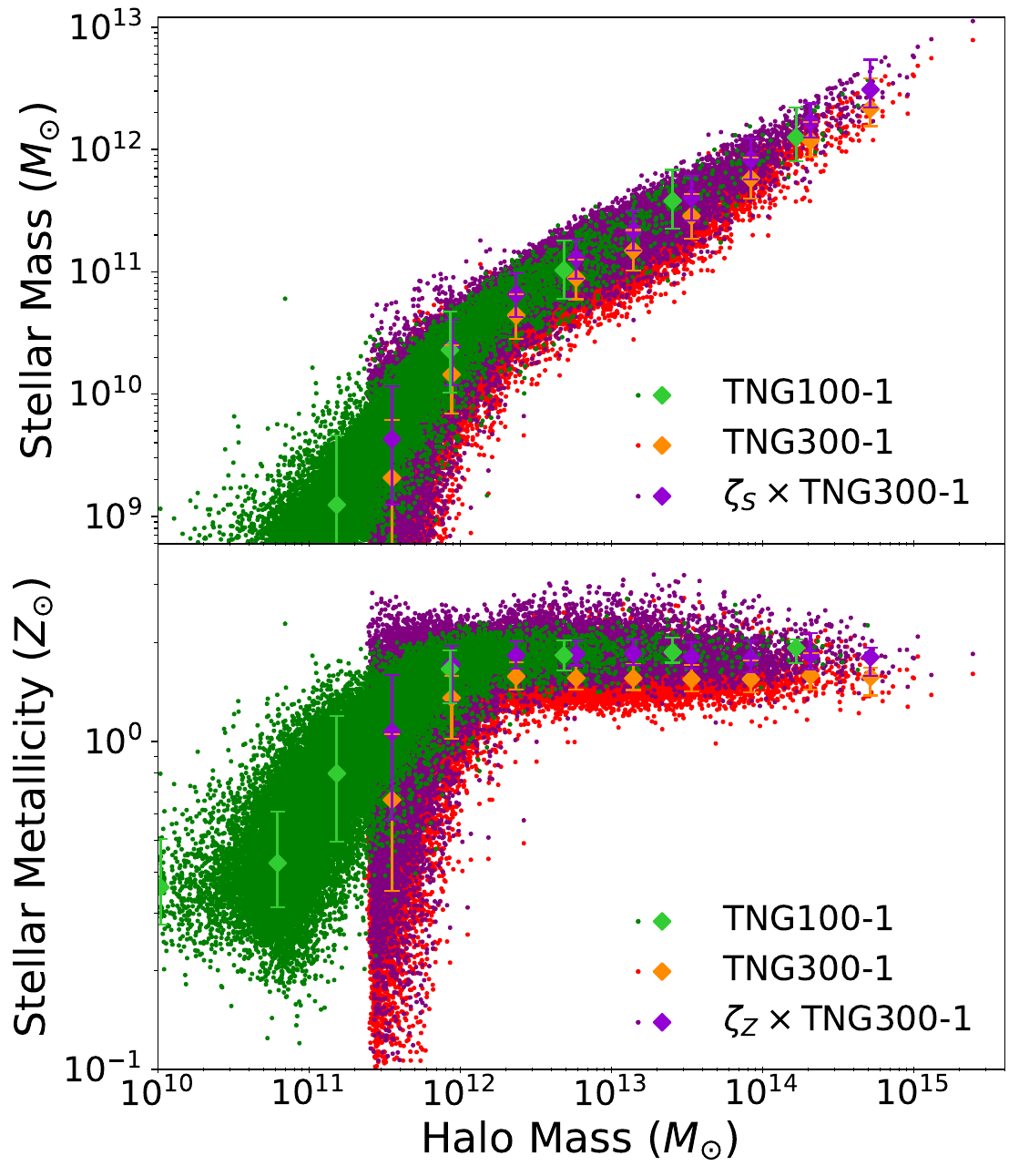}
\vspace*{-20pt}
\caption{The \ac{shmr} and halo mass-metallicity relation at $z=0$, showing the TNG100-1 data aligned with the TNG300-1 distributions, following amendment using the $\zeta$ functions. For each dataset, similarly coloured errorbars indicate the median and range between $15^\text{th}$ and $85^\text{th}$ of stellar mass or metallicity in a given halo mass bin.}
\label{fig:zetacorr}
\end{figure}

\begin{figure}
\includegraphics[width=\linewidth]{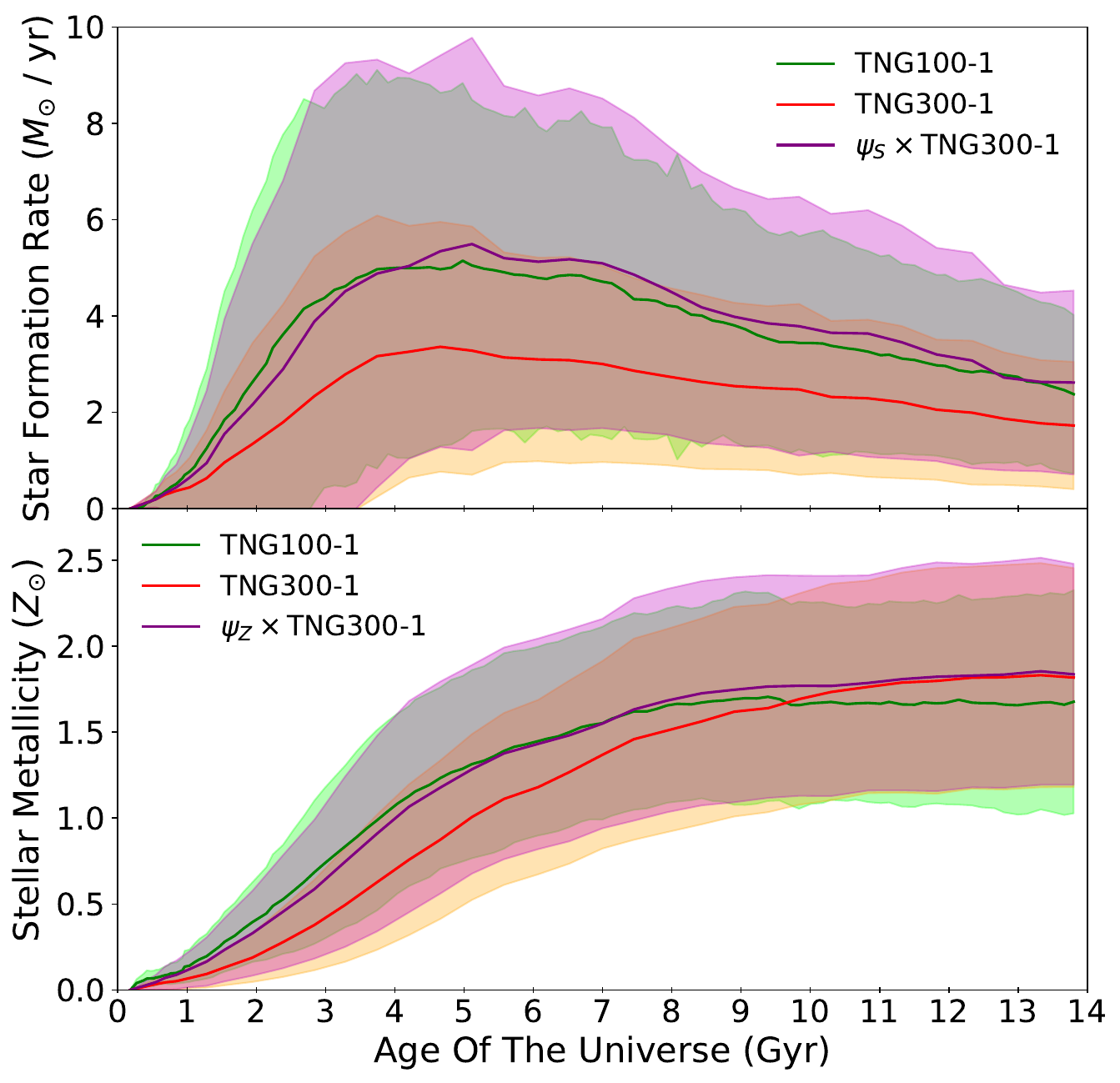}
\vspace*{-20pt}
\caption{The mean star formation and metallicity histories for central galaxies of halo mass $10^{12} M_\odot \leq M_h^{z=0} \leq 10^{12.2} M_\odot$. The standard deviation of these data as a function of time are indicated by the size of the shaded regions. This shows the TNG100-1 data aligned with the TNG300-1 means, adjusted by the $\psi$ parameters. To remove spurious features due to low numbers at early times, this figure only includes time steps with more than 100 nonzero values in all data. This applies only to times before a cosmic time of 1Gyr.}
\label{fig:psicorr}
\end{figure}

Using these corrections, we find star formation and metallicity histories in TNG300 that are accurately matched to the TNG100 data, and therefore are suitable for the neural network. These matches are displayed in \cref{fig:zetacorr,fig:psicorr}.

\subsection{Spectroscopy \& Photometry}
\label{sec:spectra}

We compute a set of spectral energy distributions from our original and predicted \ac{sfh}s and \ac{zh}s using the Python \ac{fsps} module\footnote{\href{https://github.com/dfm/python-fsps}{https://github.com/dfm/python-fsps}}. This is based on a series of \ac{ssp} spectra with \ac{imf} in accordance with the \citet{Chabrier} model. For each time step in our data, we emulate an \ac{ssp} spectrum, parameterised by the current time and metallicity of the galaxy, and weight them according to the current star formation rate. Mathematically, the full spectrum $\mathcal{F}$ is defined in terms of the \ac{ssp} spectra $f_j$ as follows:
\begin{equation}
\mathcal{F}(\lambda) = \sum_{j=1}^{N_\text{snap}} \mathcal{M}_j \ f_j(\lambda; \ \mathcal{Z}_j, \ z_j)
\label{eq:csp}
\end{equation}

Photometric magnitudes are computed from these spectra using SDSS methodology. We mimic the flux passing through the five bandpass filters used in the survey \citep{Fukugita} by integrating the spectrum over the response functions from SDSS, and compute absolute magnitudes and colours.

The distribution of colours from several combinations of bands is bimodal, where ``blue'' galaxies have ongoing star formation, ``red'' galaxies are quenched, and galaxies in the transition phase from blue to red are members of the ``green valley'' \citep{Nelson2018}. The colour-mass diagrams for centrals and satellites, inferred from network predictions are compared here with results from \ac{illtng}.

We also compute the $H\alpha$ luminosities of our galaxies by taking the SFH-weighted sum of \ac{ssp} values, as given by \ac{fsps}:
\begin{equation}
L_\text{gal}^{H\alpha} = \sum_{j=1}^{N_\text{snap}} \mathcal{M}_j \ L_j^{H\alpha}
\label{eq:Halpha}
\end{equation}

\section{Neural Network Design \& Data Preprocessing}
\label{sec:preprocessing}

\subsection{Quantile Transformation}
\label{sec:gqt}

It is common practice, and indeed required, to normalise datasets in many applications of machine learning. Normalisation is especially important for datasets with variables of very different and large ranges, such as ours.

For much of our data, we found that a simple scaling relation was inadequate as several quantities are not represented equally. For example, halo masses are heavily over-represented at low values, while large values are under-represented in our sample, leading to poor training. A partial solution to this issue was a \ac{gqt}: an operation supported in the SciKit-Learn \citep{SKLearn} Python Library.

Quantile Transformations work as follows. Let $x_i$ be a data point in the original dataset. If the distribution of these data is normalised, one may interpolate through this and integrate it to establish a \ac{cdf} $\Psi$, which increases monotonically from 0 to 1. $\Psi$ therefore maps the datapoint $x_i$ onto its corresponding quantile value. It is therefore very simple to map this onto a second probability distribution provided that its \ac{cdf} and its inverse function are analytical. Specifically, if this second distribution has a \ac{cdf} $\Phi$, then the data $x_i$ can be mapped onto this distribution like so:
\begin{equation}
y_i = \Phi^{-1} \left( \Psi ( x_i ) \right) 
\label{eq:qt}
\end{equation}

\begin{figure}
\includegraphics[width=\linewidth]{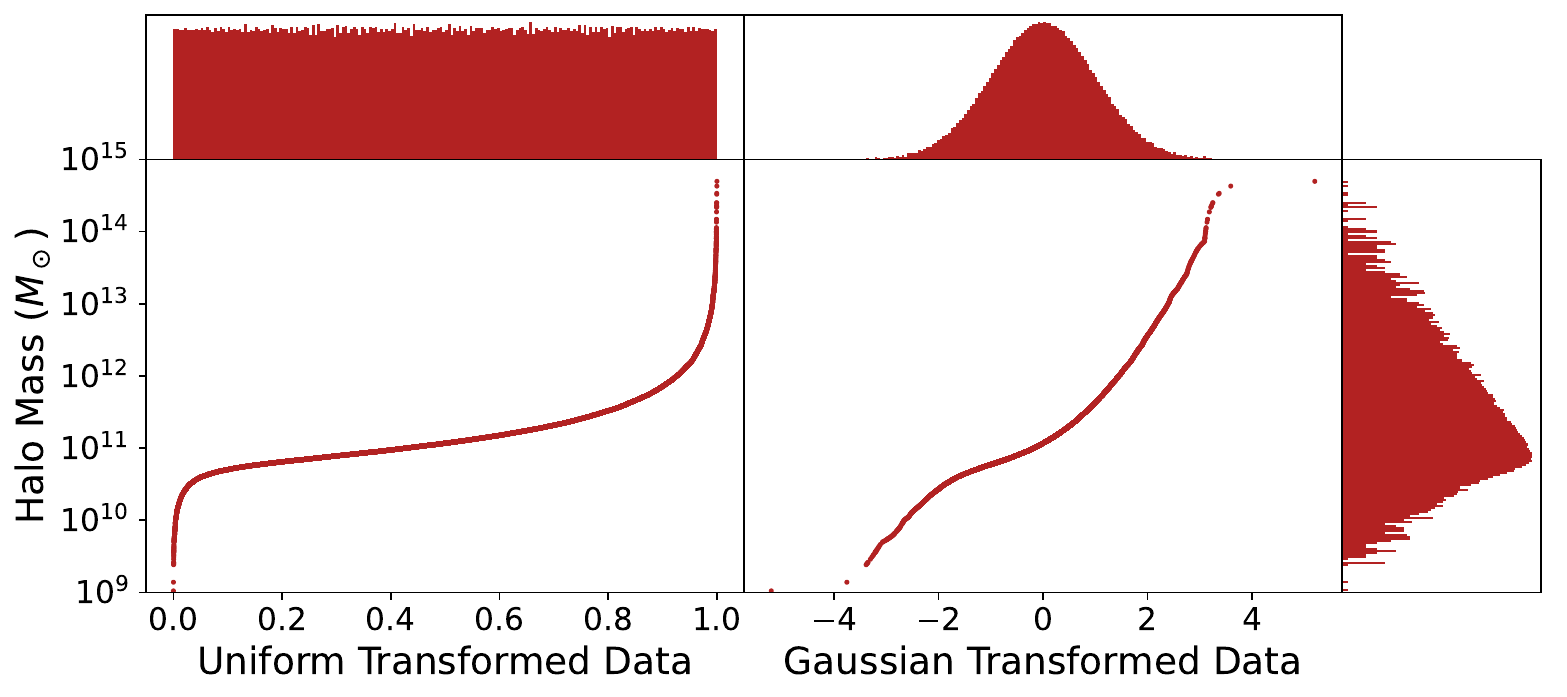}
\vspace*{-20pt}
\caption{The mapping of the TNG100-1 set of halo masses (vertical axis) to a uniform distribution (left) and Gaussian distribution (right). Histograms of these datapoints are shown along every axis, including the halo mass distribution on the vertical axis. This graphic shows that a large range of data, namely halo masses above $10^{12.5} M_\odot$, corresponds to a very narrow range in the uniform distribution, which results in high sensitivity to small differences in the transformed data. This makes a uniform distribution unsuitable for making predictions from our data, and this is why we have chosen to transform to a Gaussian distribution instead.}
\label{fig:qt}
\end{figure}

If we intend for the $y_i$ to be Gaussian distributed, then \cref{eq:qt} takes the specific form:
\begin{equation}
y_i = \sqrt{2} \erf^{-1} \left( 2 \Psi( x_i ) - 1 \right)
\label{eq:gqt}
\end{equation}

and can be inverted when returning predicted data to physical values:
\begin{equation}
x_i = \Psi^{-1} \left( \frac{1 + \erf \left( \sfrac{\mathlarger{y_i}}{\sqrt{2}} \right)}{2} \right)
\label{eq:gqtinv}
\end{equation}

A \ac{gqt} was chosen specifically because the domain of our transformed data resides entirely between 5 and -5, making it suitably normalised for the neural network. SciKit-Learn also offer a transformation to a uniform distribution, however the Gaussian transformation proves much more suitable for our data. \Cref{fig:qt} shows how our TNG100-1 sample of logarithmic halo masses is mapped onto these distributions. In the uniform case, a number of data points are strongly skewed towards the edges of the distribution. This narrow margin containing such a broad range of values means that there can be a large error in the true data when making predictions based on a uniform distribution.

\subsection{Vector \& Scalar Normalisation}
\label{sec:vecscanorm}

\begin{figure}
\includegraphics[width=\linewidth]{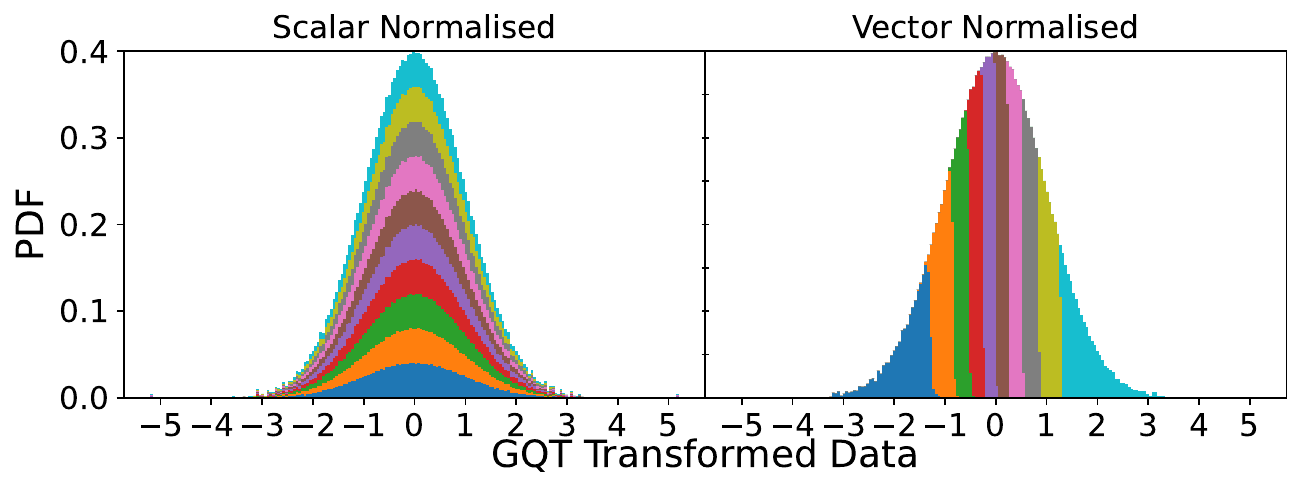}
\vspace*{-20pt}
\caption{A depiction of the distribution of arbitrary, monotonic data after applying scalar (left) and vector (right) \ac{gqt} normalisation. Data points at successive time steps are shown in different colours. All time steps share the same normal distribution when transformed independently of other time steps, i.e. when scalar normalised. When vector normalised, the data points are transformed according to the full range of the quantity's value over time, hence each time step's distribution is relative to another. Stacking the full set of time step histograms results in the Gaussian distribution of the full dataset, regardless of normalisation method.}
\label{fig:vecscanorm}
\end{figure}

In the context of time-dependent variables, the \ac{gqt} can take two specific forms. In one form, we consider that the behaviour of a dark matter quantity as a function of time can influence the present-day state of the galaxy, and so we intend to preserve the geometry (or shape) of the property's history. In the other, there may be physical or systematic differences between time steps which imply that there is no physical meaning to the gradient of the variable with time, and instead the absolute value at a given point in time is more significant. We introduce two forms of normalisation: vector and scalar normalisation, to incorporate time-dependent properties differently. The effect that each normalisation has on temporal data is illustrated in \cref{fig:vecscanorm}.

With vector normalisation, the \ac{gqt} is fit to the variable at all time steps simultaneously. The transformed variable corresponds to the original variable's value regardless of the time at which it is defined. \Cref{eq:gqt} refers to a \ac{gqt} for a time-independent variable. If $\Psi$ is the \ac{cdf} of $x_i$: the 1D set of values of the temporal variable, the transformed variable as a function of time is defined by applying the transformation to the original, 2D dataset:
\begin{equation}
y_i(t_j) = \sqrt{2} \erf^{-1} \left( 2 \Psi( x_i(t_j) ) - 1 \right)
\label{eq:vecnorm}
\end{equation}

where $t_j$ is a specific time step. These are implemented into the recurrent input layer as a time-dependent vector, whose normalised value serves as an absolute indicator of the original data value, and therefore the temporal gradients of the variable are preserved.

Any variable for which the \ac{gqt} is fit with no respect to time is considered scalar normalised. Here there exists a separate \ac{cdf} for the variable at each time step, and thus a list of independent sets of transformed variables, separated in time:
\begin{equation}
y_i^k = \sqrt{2} \erf^{-1} \left( 2 \Psi_k( x_i^k ) - 1 \right)
\label{eq:scanorm}
\end{equation}

\subsection{Neural Network Architecture}
\label{sec:network}

We have developed two semi-recurrent neural networks in TensorFlow \citep{TensorFlow}, which have been trained separately on central and satellite galaxy data from \ac{illtng}. In each network, the temporal component of the halo data constitutes the recurrent input data, while the remaining zero-redshift quantities such as halo mass are processed in the second, dense input layer.

While \ac{illtng} contains 100 snapshots in time, for the sake of reducing the complexity of the neural network for better convergence, we evaluate all time-dependent halo and galaxy properties with a step size of three snapshots, thus computing a 33-element vector.

\begin{figure*}
\includegraphics[width=\linewidth]{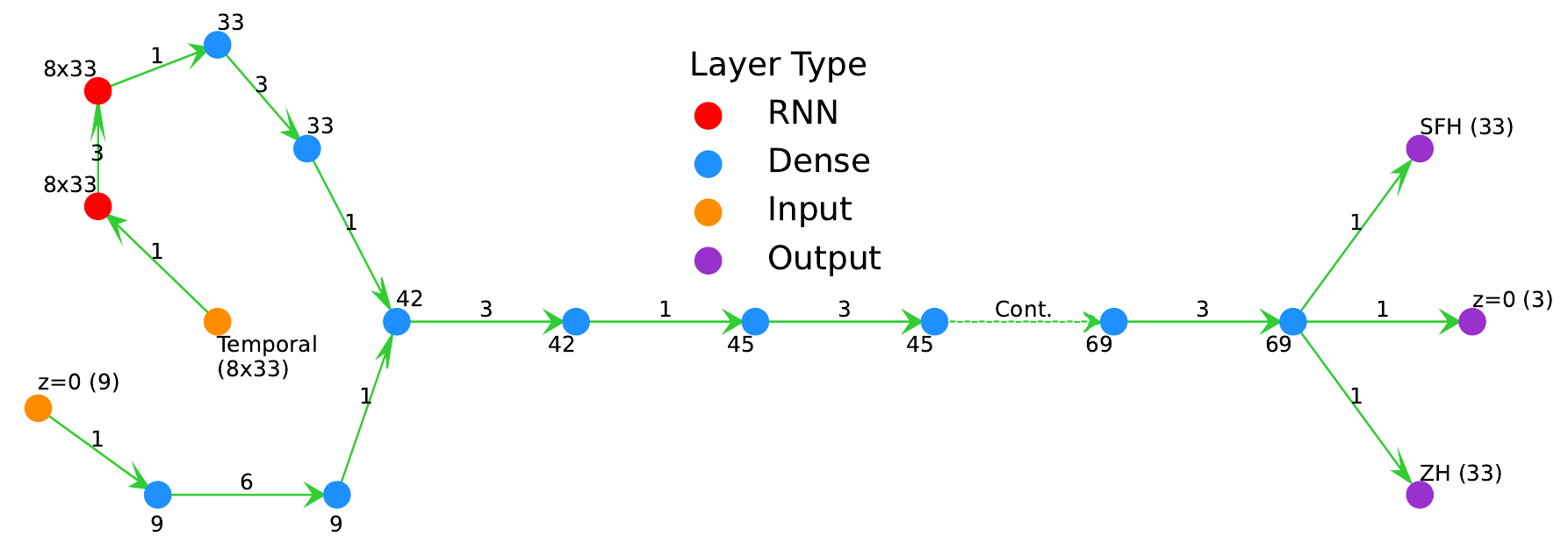}
\vspace*{-20pt}
\caption{An illustration of the architecture of the neural network for central galaxies. Each point in this diagram represents a fully connected layer, whose dimensionality is indicated by a numerical label; with the exception of purple points, which represent a subset of the final, 1D output layer. An input layer exists for the time-dependent and time-independent halo properties each, whose outputs are conjoined at a 42-node dense layer. The temporal input layer and recurrent layers are two-dimensional, comprising eight variables defined over 33 time steps. Each arrow indicates a connection between consecutive layers, while the arrow's label indicates the number of times that this connection repeats (i.e. ``3'' indicates that four consecutive hidden layers exist to accommodate three such connections). The dashed line arrow indicates that the process of adding three nodes to every fourth hidden layer repeats up to the point that there are 69 nodes per layer. The final set of connections returns the baryonic data output, consisting of star formation and metallicity histories, and zero-redshift galaxy properties.}
\label{fig:nncent}
\end{figure*}

The architecture for the central neural network is depicted in \cref{fig:nncent}. A sequence of number-labelled points indicate layers of different types and their dimensions. Each of our networks begin with a two-dimensional input of features defined at 33 time steps, and a one-dimensional set of scalar features. After multiple dense or recurrent layers, the one-dimensional layers which succeed them are eventually concatenated, allowing both sets of data to predict baryonic properties. The recurrent and dense input quantities are discussed at length in \cref{sec:timedependent,sec:timeindependent}, respectively.

The satellite neural network is identical in design to the central network shown in \cref{fig:nncent}, except that there exist seven temporal and eleven non-temporal input quantities, which again are discussed in \cref{sec:timedependent,sec:timeindependent}. The input layers therefore have dimensions of 11 and 33x7 nodes, combining in the same sequence to make a 44-node dense layer instead of 42. This is followed by a set of 45-layer nodes, and the rest of the satellite network remains true to the central design.

Two separate models exist due to the inclusion of variables which are not defined for central subhalos, such as the time of infall into a larger host halo. Satellite galaxies also have different evolutionary properties such as decoupled growth of the subhalo and host halo, which are included in this model, and typically have larger stellar mass fractions and scatter in stellar mass at high halo mass when compared with centrals \citep{Engler}, and these differences may not have been distinguished in a composite model.

Due to the sparse nature of input quantities such as halo masses and overdensities, gradient saturation renders the network inoperable when using highly nonlinear activation functions such as tanh or sigmoid. When trialling the \ac{relu} activation function, defined as follows:
\begin{equation}
\text{ReLU} ( x ) \ \equiv \ \text{max} \left[ 0, x \right]
\label{eq:relu}
\end{equation}

we find no noticeable saturation effects. However, this network was subject to the Dying \ac{relu} Problem \citep{DReLU}, in which the gradient for any \ac{relu}-activated node is automatically zero if its input is negative, thus all subsequent iterations from this node are zero, and will no longer contribute to training the model. To reduce the number of affected nodes, a common approach is to use the similar \ac{lrelu} activation function:
\begin{equation}
\text{L-ReLU} ( x ) \ \equiv \ \text{max} \left[ \alpha x, x \right]; \ \alpha \in (0,1)
\label{eq:lrelu}
\end{equation}

However, the discontinuity in the gradient of \ac{lrelu} resulted in arbitrary discrepancies between similar samples in our data. Finally, the \ac{elu} activation function was tested:
\begin{align}
\text{ELU} ( x ) \ \equiv \ \begin{cases} x &\mbox{if } x \geq 0 \\ \alpha \left( \exp (x) - 1 \right) &\mbox{if } x < 0 \end{cases}
\label{eq:elu}
\end{align}

which, with an $\alpha$ value of 1, offered a solution to both of these problems.

\ac{elu} activated networks are useful for scaling a layer's output to a near-zero mean and near-unity standard deviation, however this behaviour is only stable for sequential architectures of standardised inputs and initial kernel weights \citep{ELU, Geron}. While \ac{elu} alone offers significant performance over a batch-normalised \ac{relu} network \citep{Shah}, this instability along with the prospect of high gradients and saturation at highly negative inputs can limit the network's performance. The network is not fully deterministic, yet it converges adequately with \ac{elu} activations, unlike more direct \ac{relu}-derived functions.

All layers in these networks are fully connected, i.e. there exist no dropout layers or other dilution techniques in either network; the reduction of connections this introduces is what the \ac{elu} activation function was chosen to mitigate, alongside the two-sided gradient problem of \ac{lrelu}. We find that such dilation methods prove detrimental to the network's performance when using even \ac{elu} activation.

The \ac{illtng} data is shuffled and split such that one quarter of the data is used in the testing phase, where the predicted galaxy properties are compared with those from the simulation. Of the remaining 75\% used in the training phase, 20\% is used for validation, computing a separate loss function alongside the one being optimised. This loss function is the mean squared error between the true ($\{ y_i \}$) and predicted $\left( \ \{ f_i \} = f(\{ x_i \}) \ \right)$ outputs:
\begin{equation}
\text{MSE} \left( \{ y_i \}, \{ f_i \} \right) = \frac{1}{N} \sum_{i=1}^N \left( y_i - f_i \right)^2
\label{eq:mse}
\end{equation}

Finally, we find that a constant learning rate for the gradient descent process is inadequate; this initially has to be large to converge significantly, yet this leads to overshooting when approaching the loss function minimum. We use an exponentially decaying learning rate, which reduces in value at each training epoch. Specifically, for training epoch number $N$, the learning rate $\Gamma$ is defined:
\begin{equation}
\Gamma = \Gamma_0 \exp\left[-\frac{N}{N_0}\right]
\label{eq:learningrate}
\end{equation}

where the most suitable values for $N_0$ and $\Gamma_0$ were found by trial and error to be $10$ and $8 \times 10^{-4}$ respectively. As $\Gamma$ will eventually fall to an insignificant value, the training phase is terminated once $\Gamma \leq 10^{-3} \ \Gamma_0$; therefore there exists a total of 70 training epochs.

\subsection{Choice Of Architecture}
\label{sec:comparenetworks}

The adoption of our semi-recurrent network design was the result of an amalgamation of tests of the network's ability to make basic predictions consistently. The use of a recurrent layer for temporal input quantities amended the lack of convergence of a basic neural network layout, in which all input variables were passed to a solitary dense layer. A 2D input layer also allows multiple temporal quantities to be processed in a time grid; a dense network containing all of ours would be impractical to train.

The number of hidden dense or recurrent layers was decided based on the minimum number of layers required to achieve convergence. Each input layer was followed by the optimal number of dense or recurrent layers for the network to recognise their equal contribution before the two were concatenated. The remaining hidden layers were again optimal in number, i.e. the minimum for which there was no noticeable difference in the accuracy or consistency of predictions, while gradually increasing in dimensionality to meet the number of output nodes.

To show improved convergence of a simple semi-recurrent network compared with a dense network, we incorporate two simplified designs of a basic (dense) network and its semi-recurrent equivalent. As with the main models, all layers in these networks are \ac{elu} activated and contain no dropout. Each network is trained with only star formation histories as output, and halo mass accretion history as the only temporal input variable; the dense network will not converge with the hundreds of input nodes introduced by implementing multiple temporal variables. The remaining input quantities include final halo mass, starting time and maximum absolute accretion rate.

The basic dense network therefore has 36 input nodes, 33 of which contain the accretion rate at different times; and 33 output notes to accommodate the star formation history. There are 15 hidden layers from input to output, each with the same shape as the input layer.

The basic semi-recurrent network contains the accretion history in a recurrent input layer of shape (33, 1) and the remaining three quantities in a dense input layer. Following five hidden dense layers each, the two sequences merge to form a single 36-node layer, which is followed by 15 layers of the same shape before reaching the output layer.

The choice of number of layers in these models was decided as the optimum number of layers, as with the main model. Therefore this test of network design illustrates the difference in predictions of the simplest possible complete models. We use these two models to illustrate the advantage of the semi-recurrent design in \cref{sec:sfh2}.

\section{Predictions}
\label{sec:predictions}

This section concerns the results of the neural network models for central and for satellite galaxies, comprising the properties of the simulated galaxy formation histories and a discussion of the physical consequences on their level of accuracy.

\subsection{Galaxy Properties}
\label{sec:directpred}

\subsubsection{Stellar-Halo Mass Relation}
\label{sec:shmr}

\begin{figure*}
\includegraphics[width=\linewidth]{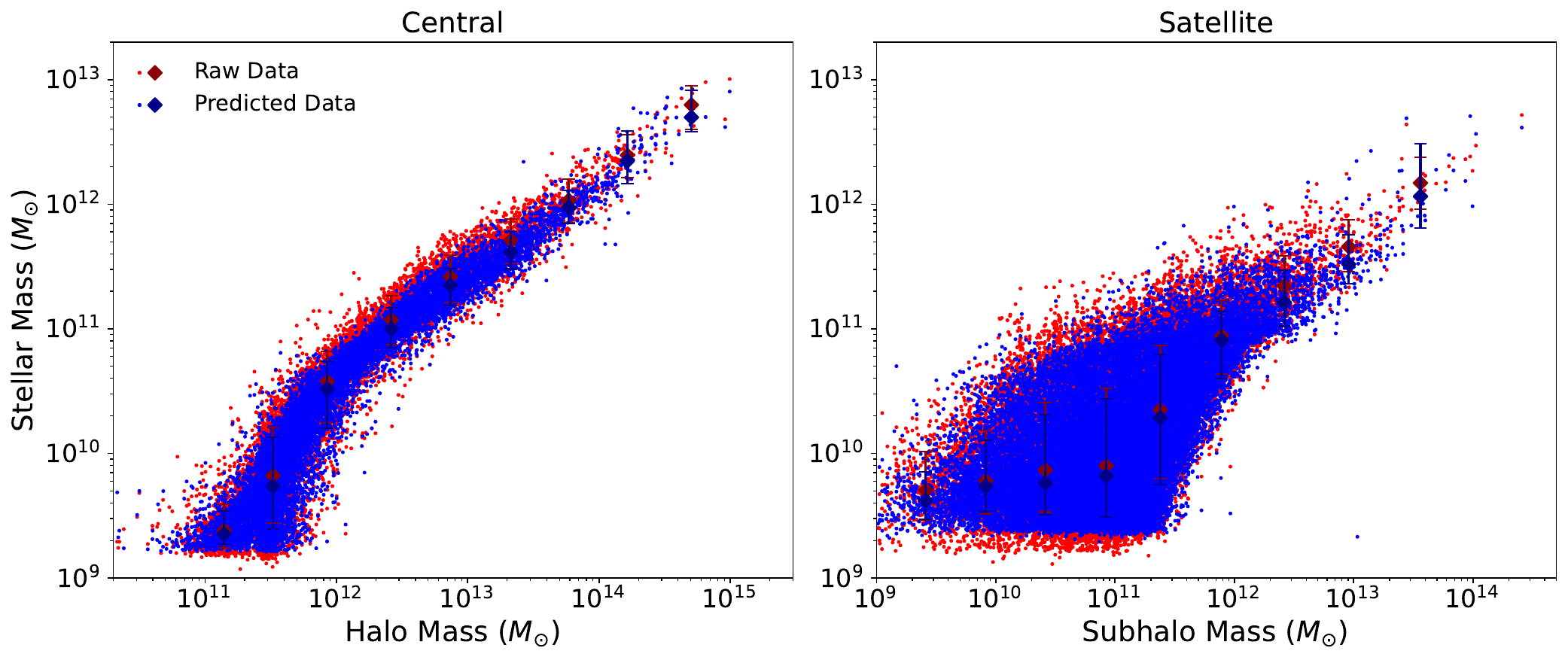}
\vspace*{-20pt}
\caption{The numerical \ac{shmr} evaluated with the true and predicted star formation rates, for central galaxies (left) and satellite galaxies (right). This is shown as a function of \ac{fof} halo mass for the former, and subhalo mass for the latter. Datapoints from the original dataset are shown in red, predictions in blue, while red and blue errorbars show the median and $15^\text{th}$ and $85^\text{th}$ percentiles of stellar mass in a given halo mass bin. The strong similarity of the shapes of these relations indicates a very good overall prediction of the star formation histories.}
\label{fig:shmr}
\end{figure*}

Numerical integration of the true and predicted \ac{sfh} for each galaxy is used to plot the \ac{shmr}, shown in \cref{fig:shmr}. Median absolute residuals in the logarithmic stellar mass are 0.079 dex for central galaxies, and 0.094 dex for satellites. 

Despite the accuracy of this result, the network under-predicts the stellar mass of most galaxies. Another prevailing problem, discussed below, is the failure of the neural network to predict short bursts of star formation, which, despite their brevity, will have significant influence on the spectroscopy of a galaxy. In general, this serves to reduce the scatter in the \ac{shmr}, and consequently this is also true for scatter in metallicity.

\subsubsection{Star Formation History}
\label{sec:sfh2}

\begin{figure*}
\includegraphics[width=\linewidth]{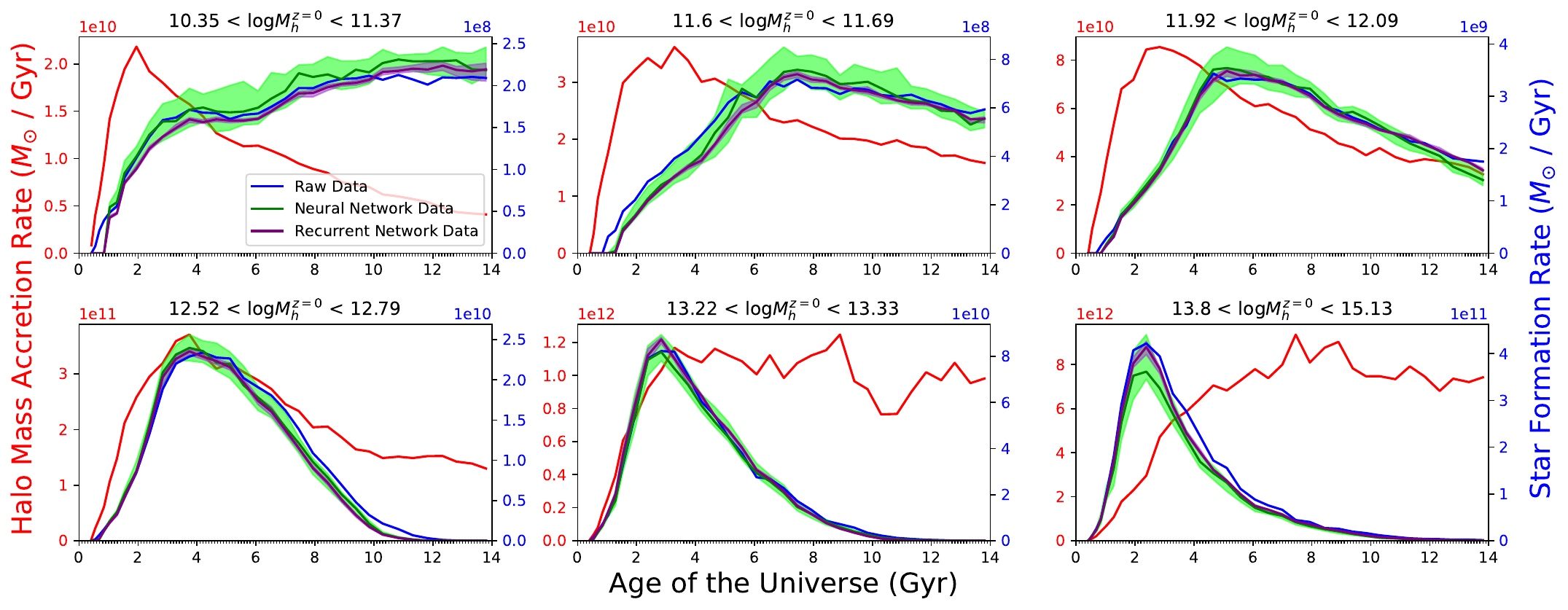}
\vspace*{-20pt}
\caption{The shape of the mean star formation history in six bins of halo mass for the central galaxy data, where predictions from a basic dense neural network (green) are compared with those from our semi-recurrent architecture (purple). Their lines and shaded regions depict the median and interquartile ranges of the predicted mean from ten independent runs of each network. These are shown alongside the true mean star formation history (blue) and the mean mass accretion rate (red). The differences in interquartile ranges between predictions of the two networks shows a significant improvement in the converging power of the network when implementing the semi-recurrent design.}
\label{fig:nncomp}
\end{figure*}

\Cref{fig:nncomp} shows the mean of the true and predicted SFHs from the simplified models described in \cref{sec:comparenetworks}, in six bins of halo mass. The general behaviour with halo mass is clearly predicted, with galaxies in higher mass halos forming their stars earlier - i.e., these networks reproduce galaxy downsizing trends with recent star-formation shifting towards low-mass halos. \par The difference in the precision of the two networks is apparent: in each halo mass bin, the mean \ac{sfh} from the semi-recurrent network is better constrained, and in most cases matches the true mean more consistently. Comparing predictions from ten independent runs, the variance in the predicted \ac{sfh} at any snapshot was typically 1-2 orders of magnitude larger in a basic dense network than its semi-recurrent equivalent. The improved precision justifies our use of a recurrent treatment of historical dark matter properties.

\begin{figure*}
\includegraphics[width=\linewidth]{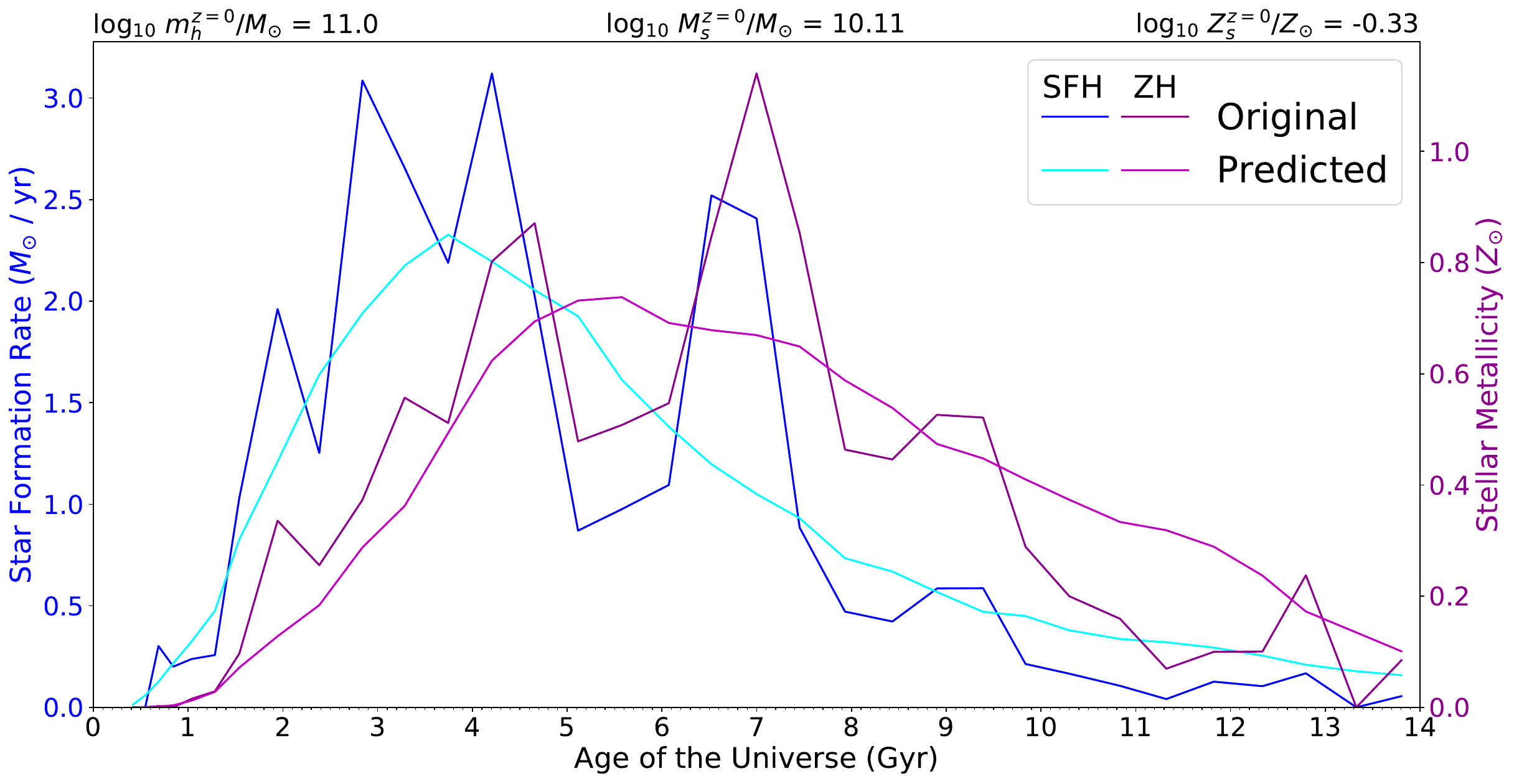}
\vspace*{-20pt}
\caption{An example formation history for a satellite galaxy of intermediate mass. The true star formation history (blue) and the corresponding prediction (cyan) are shown alongside the true stellar metallicity history (purple) and the predicted metallicity history (magenta). Numerical estimates of the subhalo and predicted stellar mass and metallicity are shown above the graph. This sample shows a modest fit to the shapes of the SFH and ZH, but not the variations on short timescales.}
\label{fig:sfh}
\end{figure*}

\begin{figure}
\includegraphics[width=\linewidth]{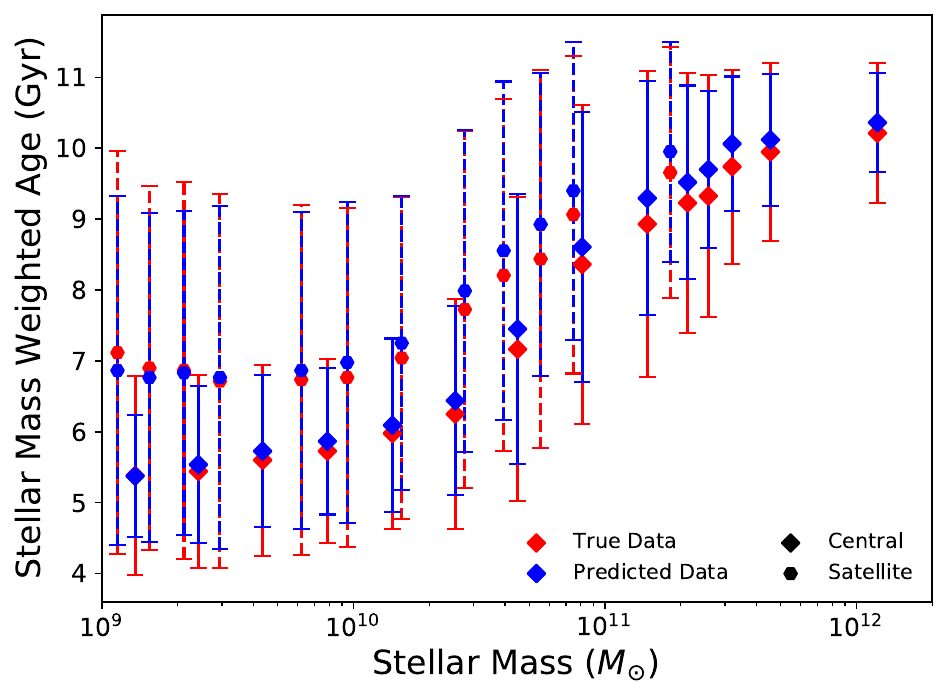}
\vspace*{-20pt}
\caption{Stellar mass weighted age as a function of stellar mass, for central (solid, diamond) and satellite galaxies (dashed, hexagon). The plot points and error bars represent the median and interquartile range of mass weighted ages in each bin. The ages derived from the predicted stellar mass assembly history (blue) are overlaid upon the values computed from the original data, showing that the general trend of ages with respect to masses are well-matched, however the offsets and reduced ranges of certain error bars are indicative of the prevalence of samples with considerable differences in mass assembly geometry.}
\label{fig:mwa}
\end{figure}

For the main network, \cref{fig:sfh} shows the predicted and true SFH and ZH of a single galaxy in more detail. The broad shape of the SFH is well recovered, as is the stellar mass, but the NN is unable to predict the intrinsic variability of the order of ~1Gyr in the original data\footnote{The true variability will of course have components at much shorter timescales, but is re-sampled here to 33 bins as described in \cref{sec:network}.}. The overall accuracy of the predicted geometries is visualised in \cref{fig:mwa}, where we show the \ac{mwa} of galaxies as a function of halo mass, for centrals and satellites separately. Here, it is easier to discern a tendency of the NN to over predict \ac{mwa}s, and to under-predict their scatter. 

We find that these are general intrinsic shortcomings of the NN. The power in fluctuations on timescales lower than around $0.1$ Gyr$^{-1}$ are suppressed in the predicted data, which can explain the lower scatter in the MWAs at fixed halo mass. We will see later that this suppression is also correlated with discrepancies between predicted and true SEDs, and propose a possible way to overcome it.

\begin{figure}
\includegraphics[width=\linewidth]{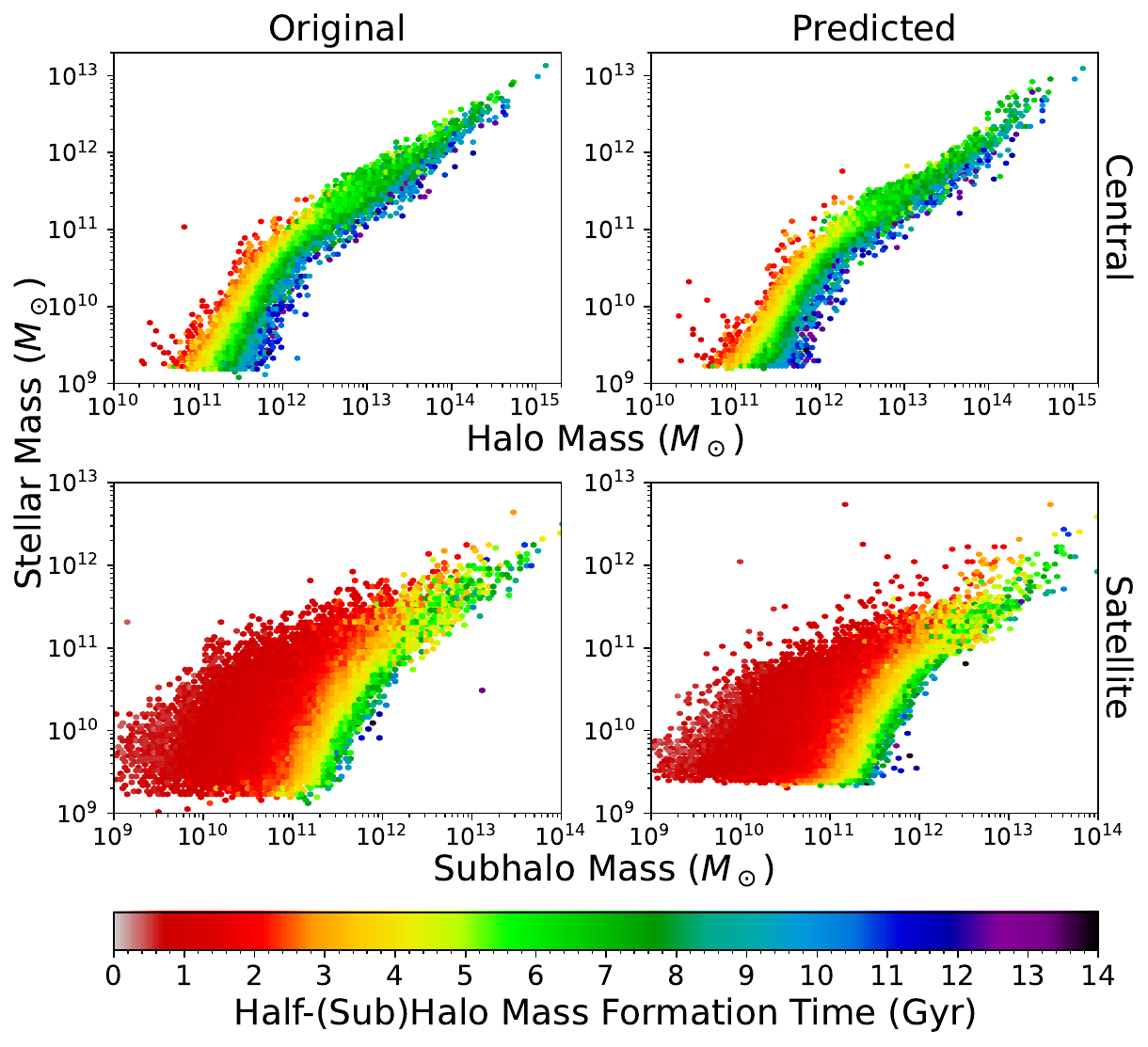}
\vspace*{-20pt}
\caption{Stellar-halo mass relations for central galaxies (top panels) and satellites (bottom panels), in 2D bins coloured according to the mean cosmic time per bin at which half of the final mass of the halo (central) or subhalo (satellite) is formed. Each SHMR is shown separately for the original TNG data and the networks' predictions by plotting according to respective stellar masses. The same trend of this property along the SHMR in both circumstances shows that this relationship is reflected in the network's predictions. Note that the low occupancy of bins on the edges of the SHMRs are sensitive to minor differences in scatter, which results in an apparent distortion of the shape of the SHMR, which is misleading.}
\label{fig:shmrs}
\end{figure}

In \cref{fig:shmrs} we show how the time at which half the final (sub)halo mass is formed varies according to the \ac{shmr} for centrals and satellite galaxies, in both original and predicted datasets. This is an important aspect of the \ac{shmr} which illustrates the relation of speed of halo growth to stellar mass \citep{Cui}. While this quantity is not an explicit feature of the neural networks, its relationship with the \ac{shmr} is preserved in the predictions of the neural network. This shows that the causal dependence of halo growth on star formation is captured by the neural network.

\subsubsection{Chemical Enrichment History}
\label{sec:zh2}

\begin{figure*}
\includegraphics[width=\linewidth]{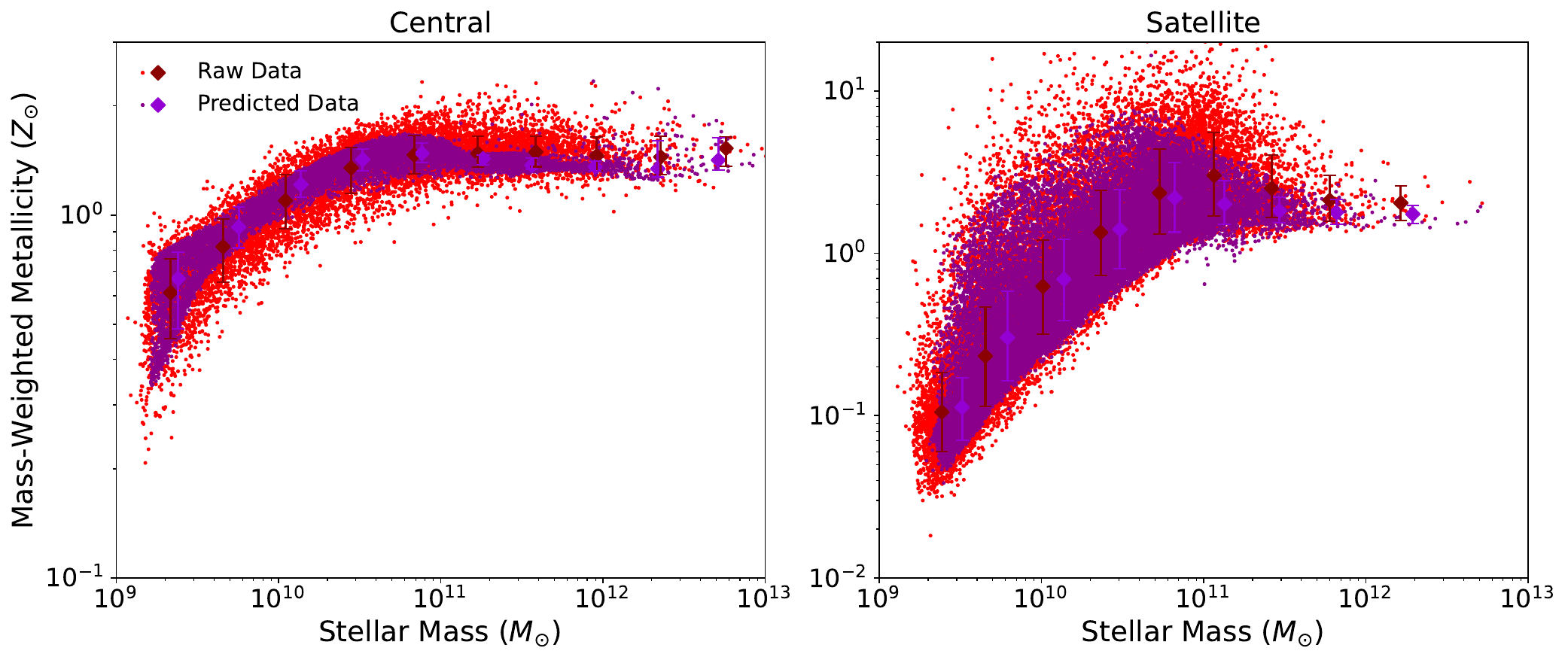}
\vspace*{-20pt}
\caption{The numerical \ac{mzr} evaluated with the true and predicted chemical enrichment histories, for central galaxies (left) and satellite galaxies (right). The true and predicted metallicities are respectively shown as a function of true and predicted numerical stellar mass. Datapoints from the original dataset are shown in red, predictions in purple, while red and purple errorbars show the median and $15^\text{th}$ and $85^\text{th}$ percentiles of stellar metallicity in a given stellar mass bin. Like \cref{fig:shmr}, the relation is well matched, however in this figure some extremities in scatter are unreached.}
\label{fig:mzr}
\end{figure*}

\Cref{fig:mzr} depicts the relation between halo mass and mass-weighted metallicity of each galaxy. The scatter in metallicity is underestimated by the neural network, to a lesser extent for satellite galaxies. Metal-rich objects in particular have lower metallicities across stellar masses. We also see a lack of high frequency information in metallicity histories such as the example in \cref{fig:sfh}.

We argue that this overall lack of high frequency information of star formation and metallicity histories is related to reduced scatter in both physical and observational summary statistics. This is discussed in \cref{sec:lumfourier}.

\subsection{Observational Data}
\label{sec:observables}

\subsubsection{Spectra \& Line Emission}
\label{sec:specsandlines}

\begin{figure*}
\includegraphics[width=\linewidth]{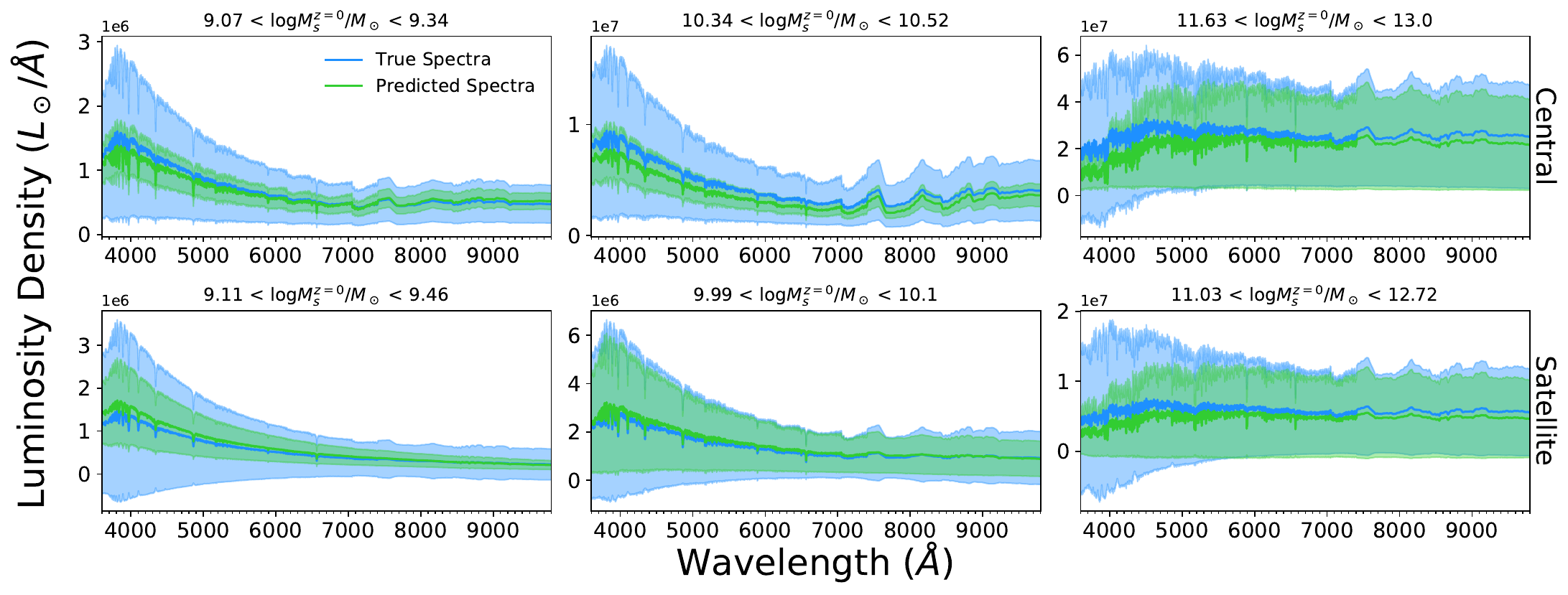}
\vspace*{-20pt}
\caption{The mean and standard deviation for stacked central (top row) and satellite (bottom row) spectra in bins of stellar mass, shown for predicted star formation and metallicity histories in green, and \ac{illtng} data in blue. Emission lines have been omitted from these plots for clarity. In the majority of samples, the continuum is generally well recovered, and is of similar amplitude. However, for high mass objects there is a reduced variance at short wavelengths, and lower mass galaxies have a smaller variance overall. This represents a poorer prediction of central galaxy spectra, with lower mean amplitudes and smaller variance than the spectra evaluated from TNG data.}
\label{fig:spectra}
\end{figure*}

\Cref{fig:spectra} shows the mean of the stacked galaxy spectra, in bins of stellar mass, with the shaded regions indicating the standard deviation of the set. The shape and amplitude of the spectra in most bins are consistent in both the original and predicted galaxies. In high mass galaxies, the NN tends to underpredict the mean luminosity and scatter at short wavelengths, and at lower masses, it slightly overpredicts the luminosity at all wavelengths.

The spectral energy distributions for central galaxies exhibit clearly under-predicted mean and standard deviation compared with satellites. While the continuum and line emission features of these spectra tend to change in the correct way with halo mass, the mean amplitude of the predicted luminosities is often below of the true mean. This discrepancy may be attributed to stellar age and metallicity dependence on the mass-to-light ratio of an SSP \citep{GallazziBell}. We will show how this is associated with the difficulty of the NN in predicting SFHs on short timescales. While this is also a problem for satellites, satellites are more likely to be quenched, and are therefore less susceptible to shortcomings in predicting SFHs on short timescales.

Although not shown, the numerical estimates of $H\alpha$ line luminosity behave similarly. The predicted distribution of luminosities is similar to the original data, yet we consistently underestimate the scatter in any given mass bin. This result suggests that line emission luminosities can be predicted, yet are sensitive to the subtle variability in the galaxy's evolution, which serves to reduce their scatter at fixed stellar mass or star formation rate.

\subsubsection{Residual Luminosity \& Stochasticity}
\label{sec:lumfourier}

\begin{figure}
\includegraphics[width=\linewidth]{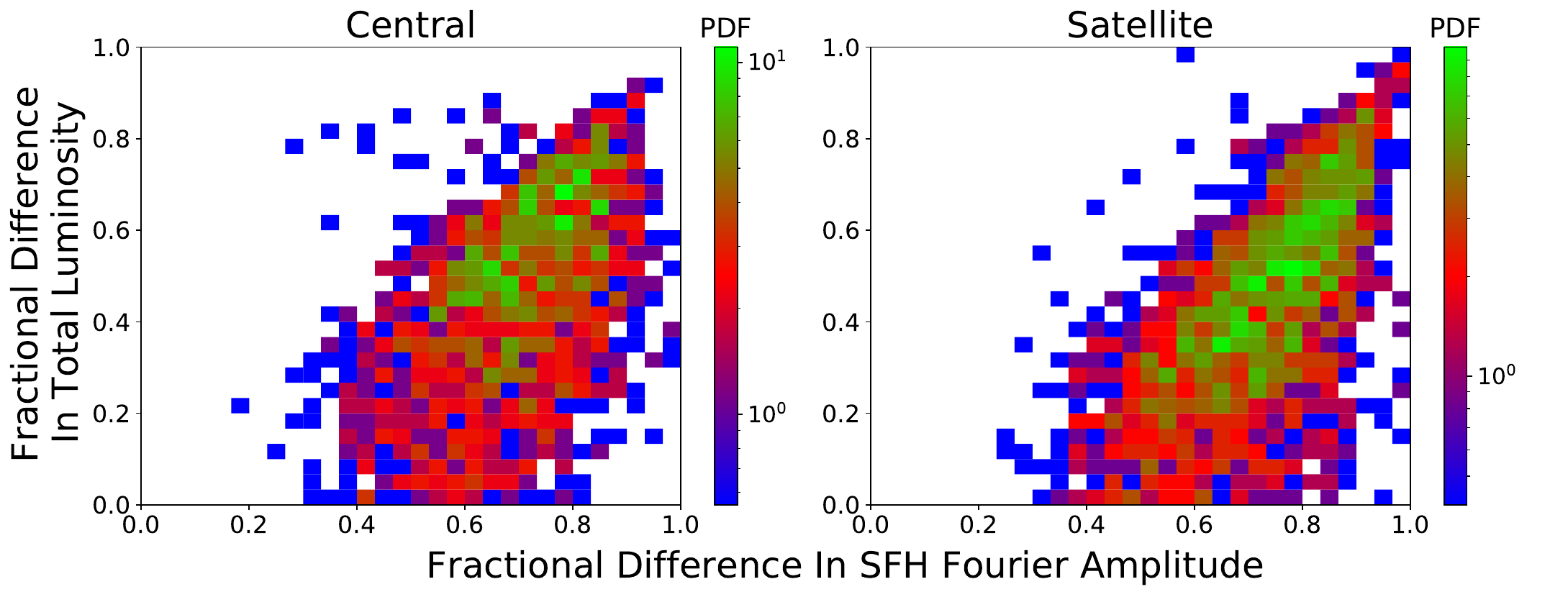}
\vspace*{-20pt}
\caption{2D histograms of the fractional difference between true and predicted total galaxy luminosity and the mean high-frequency Fourier amplitudes of their star formation histories. These are shown for galaxies between the $75^\text{th}$ and $95^\text{th}$ percentiles of $z=0$ star formation rate, and shows data within a frequency range of 0.3-1.2 $\text{Gyr}^{-1}$, i.e. a timescale range of 0.8-3.3 Gyr. This correlation between the two residuals indicates the dependence of the calculated luminosity on high-frequency star formation events.}
\label{fig:fourierlum}
\end{figure}

\begin{figure}
\includegraphics[width=\linewidth]{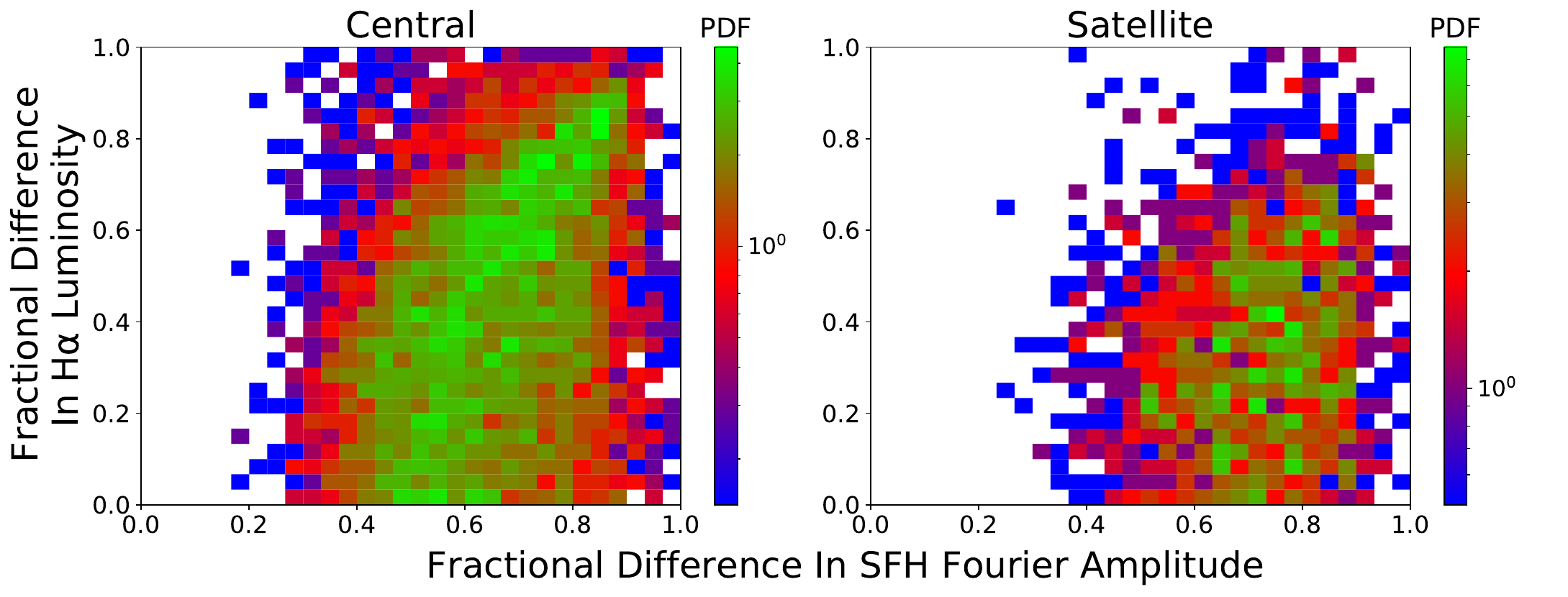}
\vspace*{-20pt}
\caption{For the same galaxies as in \cref{fig:fourierlum}, this figure shows the correlations between residuals of their high frequency SFH data and their total $H\alpha$ line luminosity. This indicates the importance of measuring short-timescale star formation events as in \cref{fig:fourierlum}, in particular at low stellar ages with the largest contribution of ionising photons.}
\label{fig:fourierha}
\end{figure}

We show the dependence of the spectra of central and satellite galaxies on short-timescale star formation events by calculating the mean of the fractional difference between the Fourier amplitudes of the true and predicted \ac{sfh}s, from $0.3 \ \text{Gyr}^{-1}$ to the Nyquist frequency of approximately $1.2 \ \text{Gyr}^{-1}$, and correlating these with the fractional difference in total luminosity in \cref{fig:fourierlum}, and $H\alpha$ line luminosity in \cref{fig:fourierha}. These are each shown for galaxies of high star formation rate at $z=0$. The frequency range of this data corresponds to variations on timescales from 0.8-3.3 Gyr.

In all of these plots, there is a clear correlation between the residuals in high-frequency Fourier modes of the \ac{sfh}s and those in their derived luminosities. This indicates that there is a significant contribution to the spectra from short star formation variability, which the networks seldom predict. This is apparent in terms of the total luminosity, which is sensitive to the size of these high frequency attributes, and the $H\alpha$ luminosity, which is sensitive to recent star formation.

\subsubsection{Photometry}
\label{sec:photometry}

\begin{figure*}
\includegraphics[width=\linewidth]{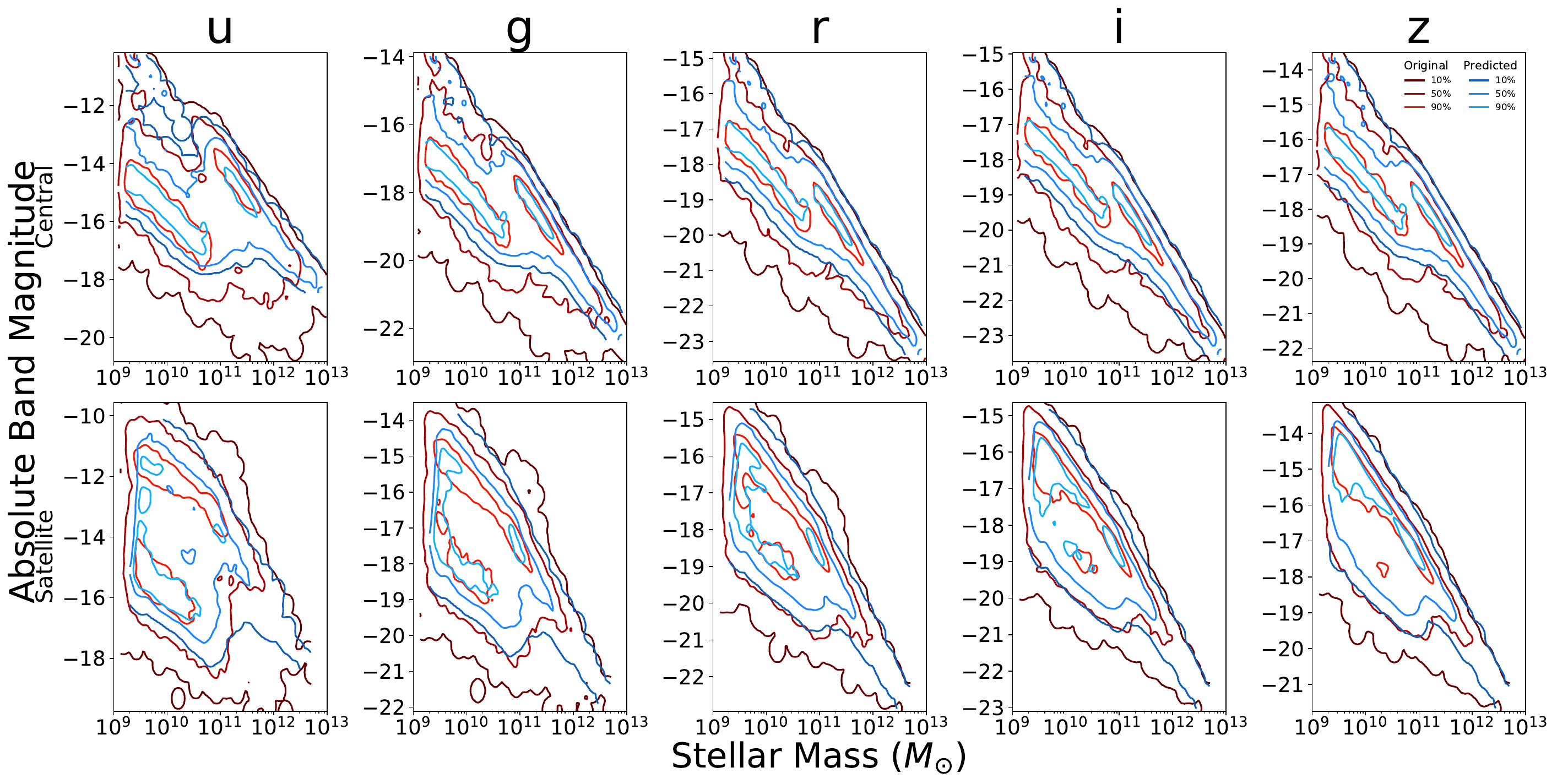}
\vspace*{-20pt}
\caption{Estimates of the five SDSS band magnitudes from the true and predicted spectra of both central and satellite galaxies, shown as a function of stellar mass, with contour lines indicating percentiles of their 2D distribution. These show a reasonable similarity in all bands despite a slight reduction in the variance of magnitudes in the predicted data. In both central and satellite data, the bimodal distribution of magnitudes can be seen in relation to mass.}
\label{fig:ugriz}
\end{figure*}

\begin{figure*}
\includegraphics[width=\linewidth]{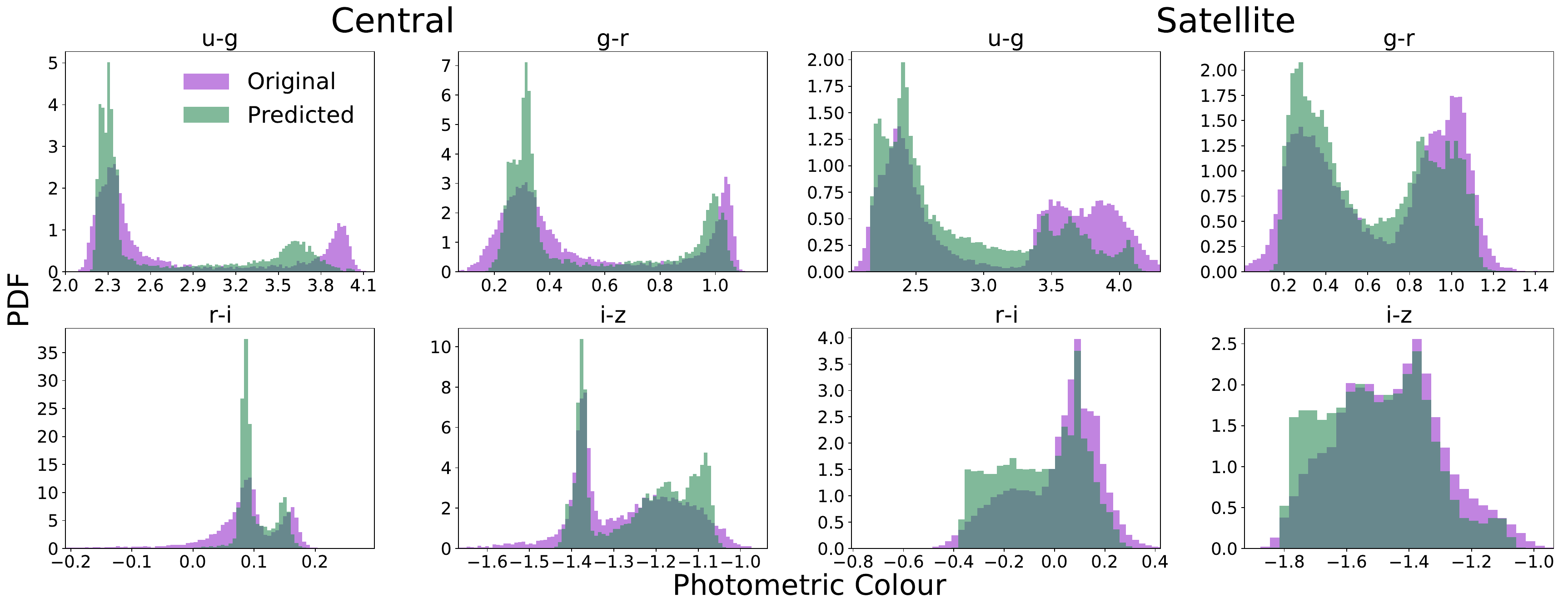}
\vspace*{-20pt}
\caption{Photometric colour distributions across the five bands, showing the differences between two consecutive bands. The distributions, mostly bimodal, are shown in purple for the original dataset, with those derived from predicted spectral energy distributions in green. We see rough agreement between datasets, however there are clear offsets in some of the data, such as bluer red galaxies in $g-r$, and significantly smaller predicted ranges.}
\label{fig:cols}
\end{figure*}

The small variance in predicted luminosities translates into a narrower range in band magnitudes. Each magnitude is plotted against the galaxy's stellar mass in \cref{fig:ugriz}, where despite the likeness of the distributions, the ``scatter'' in magnitudes is smaller, particularly for central galaxies. Colour distributions are shown in \cref{fig:cols}, depicting the expected colour bimodality in multiple band differences, for both central and satellite galaxies. The network models therefore distinguish ``blue'' star-forming galaxies and ``red'' quiescent galaxies in their predicted galaxy formation histories.

The increased difficulty in predicting the luminosity at shorter wavelengths that is shown in \cref{fig:spectra} obviously translates to systematic offsets in the bluer photometric bands. We see this in colour distributions such as $u-g$ and $g-r$, where many red galaxies are shifted towards bluer colours. Despite offsets in colours evaluated at high mass, the general inclination of galaxy colour with regard to mass is such that high mass galaxies are redder.

\begin{figure}
\includegraphics[width=\linewidth]{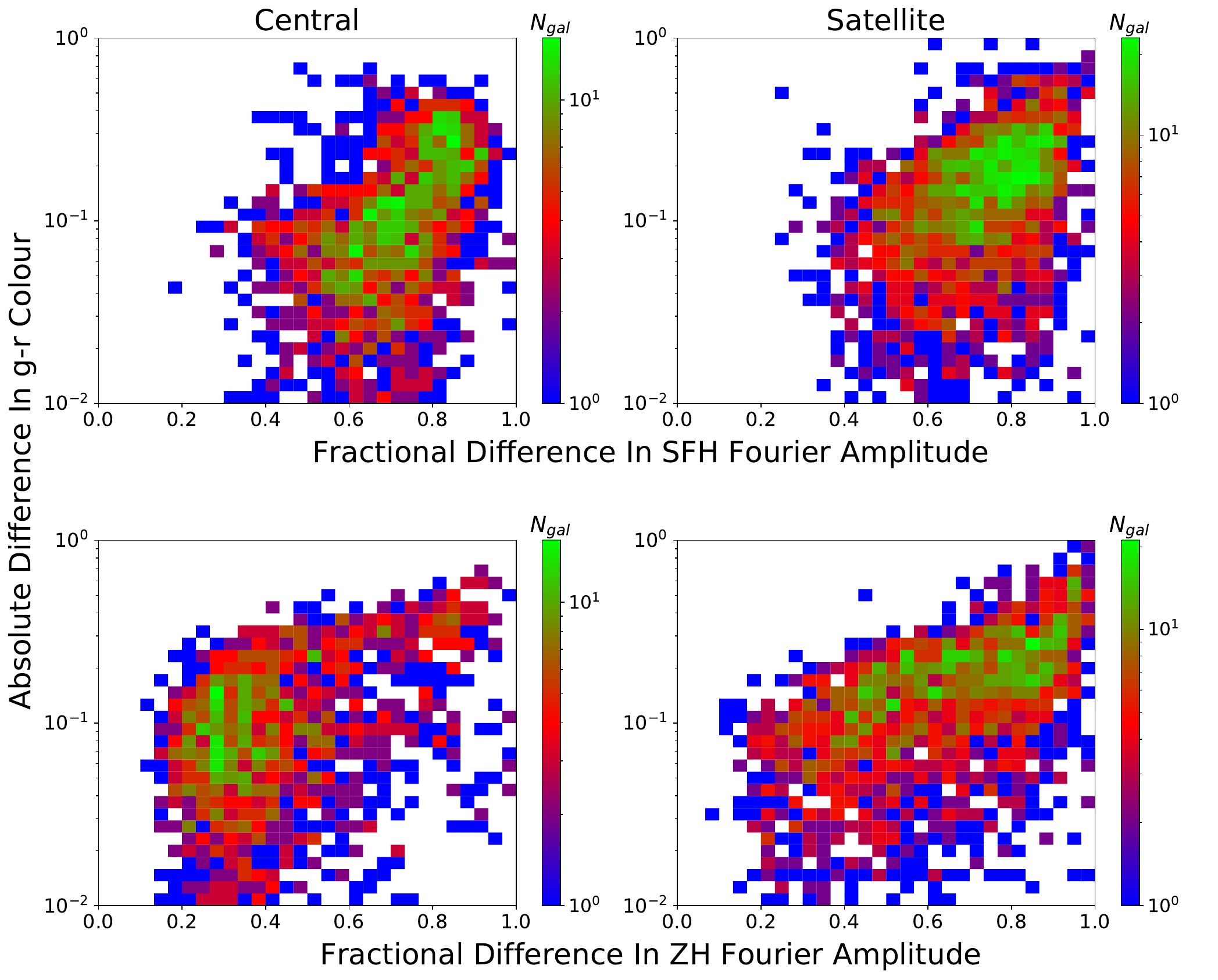}
\vspace*{-20pt}
\caption{For the data shown in \cref{fig:fourierlum,fig:fourierha}, this figure shows 2D histograms of the absolute difference between true and predicted $g-r$ colours and the mean high-frequency Fourier amplitudes of their star formation histories (top row) and metallicity histories (bottom row). This clear correlation indicates the importance of measuring short-timescale star formation and chemical enrichment events in the aim to calculate accurate colours.}
\label{fig:fouriergrcol}
\end{figure}

We showed in \cref{sec:lumfourier} that the error in the total luminosity of a given galaxy can be likened to the absence of high frequency modes in the star formation rate. It has been shown that these short star formation events impact the accuracy of photometric colours, particularly when they occur at recent times \citep{Chaves-MonteroHearin,Fraser}. We show that the error in $g-r$ colour scales with residual Fourier modes in \cref{fig:fouriergrcol}, and see that the behaviour of this correlation is similar to the effect on luminosities; but unlike the luminosity errors, this correlation is also visible for the Fourier modes in metallicity history. Noticing that some absolute colour residuals are similar in size to visible distortions of the $g-r$ distributions in \cref{fig:cols}, we see that the lack of short timescale events may explain distortions such as narrower peaks of the distribution. However, high-frequency \ac{zh} features are rare in the metal-rich galaxies whose metallicities are under-predicted, and the error in their colour may instead be due to inaccurate star formation histories.

\section{Important Features}
\label{sec:relimp}

In this section, we discuss our methods of identifying the input features of the network which have the greatest predicting power over the star formation and metallicity histories.

A commonplace metric such as a \ac{rfr} is not useful for addressing the importance of historical features; the forward-feeding influence these variables make on the final result make it difficult to compute a metric as a single number. Instead, we perform a test in which the network is trained multiple times, while groups of similar features are scrambled, in an effort to eliminate their signal.

This method is similar to permutation importance, in that we compare the performance of the model after randomising data subsets. However, it differs in the sense that the summary statistics of the disrupted model are derived from predictions, and compared with those calculated from the fiducial predictions. This allows us to measure physical properties of the data after scrambling, and identify the importance of the randomised quantities on these properties, whether or not they are explicitly given as model parameters.

\subsection{Shuffle Groups}
\label{sec:relimp2}

We evaluate the properties of the \ac{shmr}, the \ac{mhzr}, and metallicity history independently of star formation history, after randomising one of these features by replacing the training and testing data with Gaussian random noise, and training the network with this data.

There are two important aspects to this strategy. First of all, we choose these aforementioned relations in order to measure the effect on the scatter of a baryonic quantity at a fixed halo mass, thereby interpreting the effect that the shuffled quantities have on star formation and metallicity history, independently of halo mass. Second, we scramble data such that any directly related properties (e.g. $\delta_1$ and $\delta_3$) are also randomised, temporal or otherwise. Thus, each randomisation is a test of a single group of inter-related features, which we term ``shuffle groups''.

The shuffle groups to which each variable belongs are listed in \cref{tab:networks}. In summary, group 1 pertains to halo growth and contains accretion rate and final mass. Group 1a is unique to the satellite network as it contains the properties related to the satellite subhalo, while group 1 refers to its host. Group 2 contains environmental features such as overdensity and cosmic web distances, group 3 relates to dynamical properties such as circular velocity, and group 4 relates to interactive variables such as the skew.

\begin{figure}
\includegraphics[width=\linewidth]{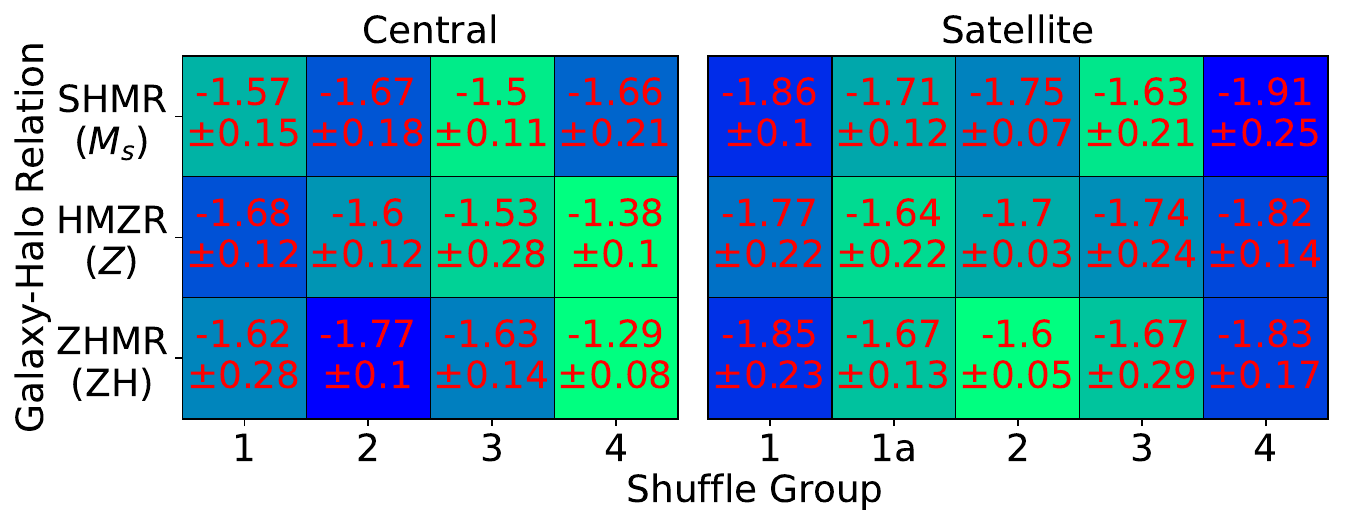}
\vspace*{-20pt}
\caption{$\Xi$ values of the stellar-halo mass relation (top row), mass-metallicity relation (middle row) and mass-metallicity history relation (bottom row), for each shuffle group in the central model (left) and satellite model (right). The text in each cell shows the median and interquartile range of $\Xi$ values obtained from ten independent runs of each network. To highlight the most significant shuffle groups, grid cells with smaller $\Xi$ values are coloured in dark blue, transitioning into bright green for higher values. Higher $\Xi$ values represent a greater importance of the given shuffle group as the model determines the given galaxy-halo relation.}
\label{fig:rt1}
\end{figure}

\subsection{Scatter Divergence}
\label{sec:relimp2a}

For ten independent runs of each randomised network, we compare the median predicted scatter to that of the un-randomised network, as a function of halo mass for central galaxies, and a function of subhalo mass for satellite galaxies. Instances where the median scatter is clearly offset from the fiducial prediction show where the scatter is supported by a property of this shuffle group.

The deviation in scatter is quantified by calculating the quantity $\Xi$:
\begin{equation}
\Xi \equiv \left\langle \log_{10} \left| 1 - \frac{\sigma_x}{\sigma_0} \right| \right\rangle
\label{eq:Xi}
\end{equation}

where $\sigma_x$ is the characteristic scatter of the given relation when shuffle group $x$ is randomised, in dex, and $\sigma_0$ is that of the fiducial result. $\Xi$ equates to $-\infty$ if $\sigma_x$ and $\sigma_0$ are identical, otherwise a larger value of $\Xi$ represents a stronger difference between scatters. The scatter is compared with the fiducial result and not the true data in order to discern the effect of scrambling features from the general inaccuracies of the model.

The angled brackets indicate that we have averaged this logarithmic ratio, over 80 loguniform bins of halo mass and weighting these bins according to their occupancy. This sample weighting is done to minimise biases from sample size limited areas, otherwise producing a misleading picture of the general size of the scatter. Additionally, $\Xi$ values can be biased towards small values if the difference in scatter frequently changes sign; thus we discard samples where the local $\Xi$ value is less than -2.

In addition to the scatter in the \ac{shmr} and \ac{mhzr}, we compute the scatter in the unweighted mean metallicity history, identifying dependencies on chemical enrichment independently of stellar mass, and thus for minority stellar populations such as late starbursts in early-type galaxies.

As well as a $\Xi$ quantity to measure differences in scatter, we apply the same method to evaluate the discrepancy in the median-filtered relations, thereby measuring the accuracy of the fit to the relations themselves. For this we replace the scatter in \cref{eq:Xi} with the logarithmic median of the given baryonic quantity. This $\Xi$ analogue is not significantly affected by any particular shuffle group, signifying no quantity with distinctive effect on the amplitudes of the three relations.

\subsection{Results}
\label{sec:relimp2b}

All values of $\Xi$ for each network, and each baryonic quantity, are tabulated for centrals and satellites in \cref{fig:rt1}, where we show the median and interquartile range of $\Xi$ values obtained from ten runs of the neural network. Values with large median $\Xi$ are considered significant, however a low interquartile range suggests that this deviation is a systematic result of the network, while higher interquartile ranges indicate poor convergence.

For central metallicity and metallicity histories, shuffle group 4, primarily involving skew, has a noticeable effect on the scatter. The skew therefore plays an important role in determining the dispersion of our metallicity histories, which may correspond to small-scale accretion events which lead to chemical enrichment. For satellites, groups 1a, 2 and 3, incorporating satellite history, overdensity and circular velocity, are the groups which affect \ac{zh} significantly, however circular velocity is less precise. One can also see notable deviations in the \ac{mhzr} scatter when scrambling groups 2 or 1a, suggesting that accretion and local environment are important to satellite enrichment.

For central galaxies, mass accretion history and circular velocity have notable offsets on the scatter of the \ac{shmr}, while circular velocity also has an effect on the satellite result. The robustness of the \ac{shmr} may be due to inherent correlations between some variables in different groups, such as the two aforementioned.

For satellites, the network performs similarly with either one of the halo and subhalo mass histories being randomised, with little change to the \ac{shmr} if only one is randomised. While the final masses and growth histories are important to the mean \ac{shmr}, their shuffle groups, containing infall parameters such as scaled infall time, have only a subtle influence on their scatter.

One can reconcile the halo growth rate and circular velocity with the \ac{shmr} scatter, as the halo's early growth and internal dynamics will quickly determine the growth of its galaxy, which restricts the feasible scope of future star formation rates. The future of its chemical enrichment can be more closely linked to environmental changes, which can be attributed to the abundance of star-forming or metal-rich gas in overdense regions, where the radial distribution would indicate the tendency for this to be accreted into the target galaxy. However, the expulsion of star forming gas owing to internal feedback mechanisms can impact the metallicity as well.

\begin{figure*}
\includegraphics[width=\linewidth]{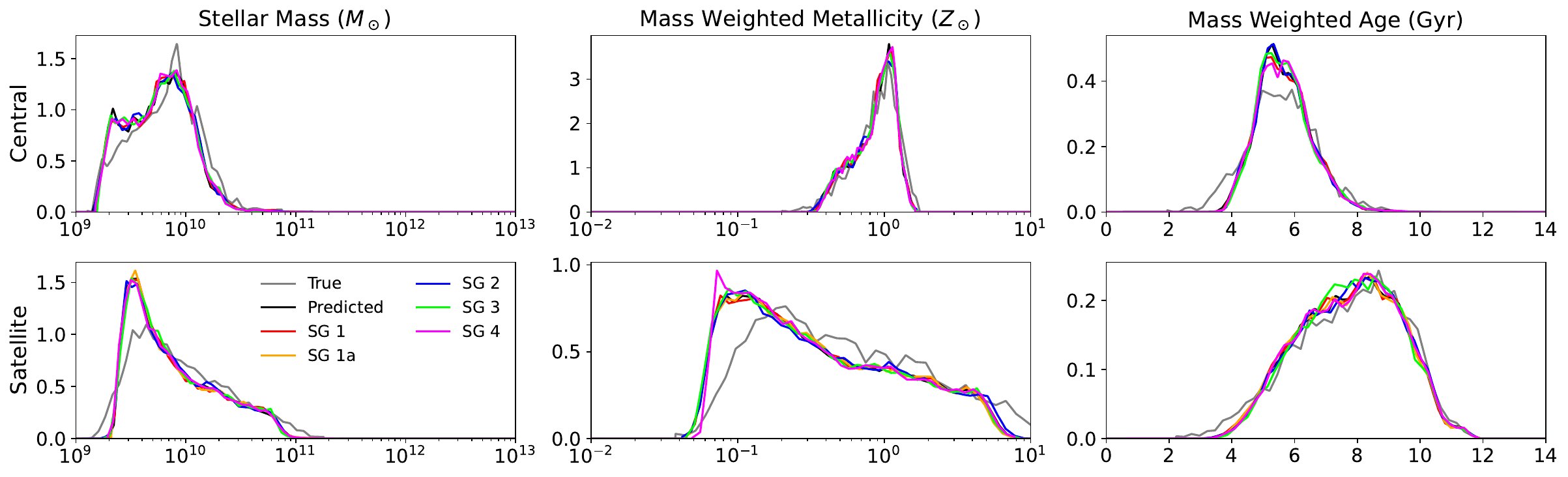}
\vspace*{-20pt}
\caption{Smoothed PDFs of stellar mass, metallicity and \ac{mwa} derived from star formation and metallicity histories in an intermediate halo/subhalo mass bin. This shows the distributions obtained from the original TNG data in grey, the median from ten fiducial predictions in black, and the median result from each randomisation in coloured lines. The distribution from each randomisation is subtle and does not show a clear difference from the best predictions, nor do they bring the network closer to the true result. If the x-axis is logarithmic, the PDF is a function of the logarithmic value.}
\label{fig:rt3}
\end{figure*}

\subsection{Median Value Divergence}
\label{sec:relimp2c}

We compare the distributions of mass, metallicity and \ac{mwa} in an intermediate halo mass bin for each network and each randomisation in \cref{fig:rt3}, shown next to the median of ten ordinary predictions, and the true data. As the network marginally underpredicts the scatter in stellar mass, and under-predicts metallicities, the true distributions are distinct from our predictions: histograms of stellar mass are marginally narrower in our predictions, and metallicity histograms are slightly offset. If the distribution from any randomisation is closer to the true distribution, it suggests that a member of this shuffle group is misleading the network, while larger differences suggest the group contains necessary information.

We find no evidence of a shuffle group which improves the network's performance if scrambled, as is the conclusion from evaluating a $\Xi$ analogue for median data values; however there is a small difference in the shape of satellite \ac{mwa} distributions when group 3 is randomised. While the scatter ratio shows the dependence of other shuffle groups in different halo mass regimes, this analysis of data in a narrow halo mass range shows that shuffle group 3 is additionally important for distinguishing similar samples. It is not explicit, however, whether this dissimilarity serves to deform the predictions, indicating that the shuffle group contains a necessary detail of the \ac{ghc}; or whether these features correlate with the \ac{illtng} distributions, which would imply that shuffle group 3 deceives the satellite network.

\subsection{Subgroup Randomising}
\label{sec:relimp3}

\begin{figure}
\includegraphics[width=\linewidth]{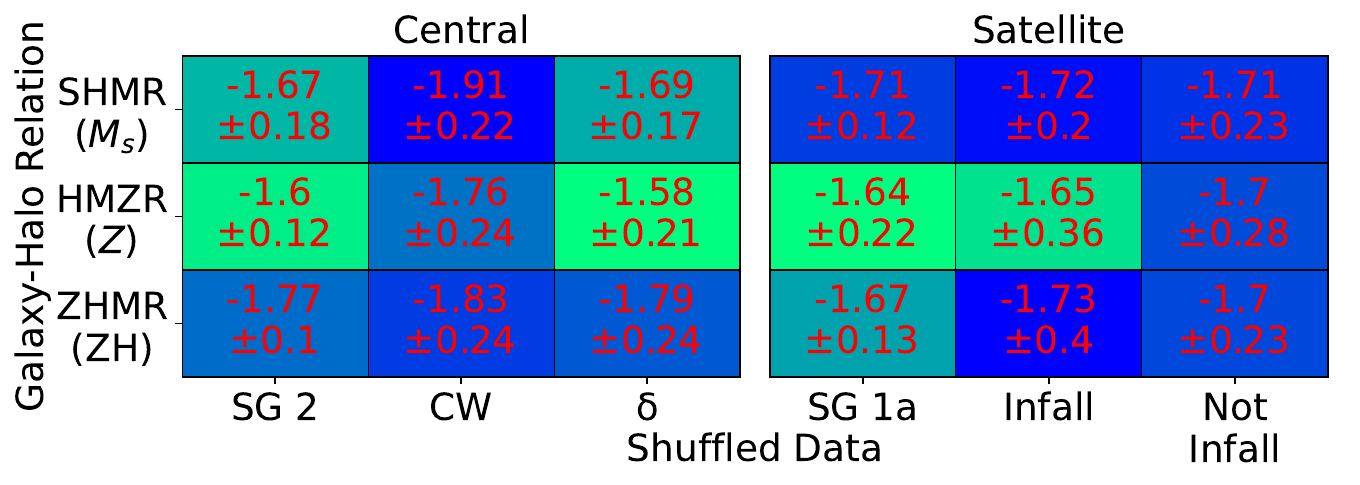}
\vspace*{-20pt}
\caption{$\Xi$ values as in \cref{fig:rt1}, instead showing the impact of shuffling certain subsets of a shuffle group. In each table, the left column shows the result for the full shuffle group as in \cref{fig:rt1}. The centre column shows the result for the data of interest, while the right column shows the result for the remainder of the shuffle group. This shows overdensity to be the dominant component of the second shuffle group for central galaxies, while infall parameters have a noticeable effect in satellite galaxies.}
\label{fig:rt2}
\end{figure}

One can scramble individual quantities or subsets of the shuffle group, which may have an important influence on the galaxy's formation history, in spite of physical correlations within the shuffle group. In \cref{fig:rt2} we show this result for two examples, one for each network, compared with the result from shuffling the remaining shuffle group members and the full shuffle group.

We test scaled infall time, formation time and infall mass ratio in the satellite network. While these may be important quantities for satellite \ac{sfh}, in \cref{fig:rt2} we see that the difference between predictions with these infall parameters and the rest of shuffle group 1a is small, suggesting that they have little effect on the performance of the satellite network. This may owe to the onset of the target's satellite phase being inferred from the growth histories of the subhalo and its host. On the contrary, the deviation from scrambling infall parameters is larger, indicating that it is useful to utilise them explicitly. The metallicity deviation is also larger for infall only, however poorly constrained, which may indicate their importance.

\citet{Donnan} indicate that star formation histories and gas phase metallicities in \ac{illtng} are modulated by the cosmic web. We randomise cosmic web distances in the central network, and compare the predicted scatter with the result when scrambling overdensities as the remaining data of shuffle group 2. Though not shown here, shuffling only the cosmic web distances has a noticeable effect on the scatter in the mass weighted metallicity of a handful of high mass galaxies. Yet in our $\Xi$ tables, showing the overall effect, the cosmic web is insignificant, and the offset is larger and applies to most galaxies when randomising overdensity. This, along with the larger interquartile ranges of $\Xi$ for the cosmic web, would suggest that the network favours the overdensity components when effectively constraining the mass weighted metallicity.

\section{Discussion}
\label{sec:discussion}

This NN model has been shown to infer the main trends of galaxies' star formation and metallicity histories with halo mass, and to broadly reproduce key observational results, such as downsizing and galaxy colour bimodality. In this section we reflect on shortcomings, insights and applications of the model.

\subsection{Baryonic Information}
\label{sec:baryons}

A possible factor limiting the predictability of the \ac{sfh} is the lack of information of gas properties, such as mass, metallicity and temperature, all of which are critical in the star formation process. Our model will predict any correlations with dark matter environment and stellar assembly for which gas properties are causally intermediate, however unconnected gas features which influence star formation may remain unrecognised. Studies show the importance of the gas content of merging halos on star formation \citep{Hani, Trevisan}; something not modelled in our analogues for merger histories.

The metallicity of the gas depends on its location in the cosmic web, and is governed largely by gas fractions and specific inflow rates, independently of halo or stellar mass \citep{Donnan, Torrey, vanLoon}. However, our cosmic web distances carry little weight in predicting the \ac{sfh} and \ac{zh}, likely due to the expectedly low sample size at fixed mass, or the fact that halo mass accretion histories and other quantities are better predictors of the shape of the \ac{sfh}. \citet{Galarraga-Espinosa} argue that the effect of cosmic web filaments on star formation either enhance or quench galaxies depending on their scale, while the number of small filaments connected to the galaxy is a more robust measure of star formation enhancement. In this work, the larger deviation seen from scrambling overdensity suggest this is a stronger constraint than the cosmic web for modelling the metallicity history of central galaxies.

Additionally, the importance of the baryonic content of merging subhalos would suggest that we may have captured star formation events more accurately if we had historical information of several of its most massive progenitors. However, designing such a network would prove difficult. Earlier attempts to characterise the merger history with a single variable in this work, such as median merger ratio per snapshot, were ineffective in improving the network's performance. A model may exist in future which predicts the baryonic growth of a halo's most massive progenitors, which by correlation with the \ac{mpb} can predict the amplitude and timescale of merger-induced starbursts. Alternatively, one may use a temporal analogue to the specific merger rates evaluated for discrete time intervals in \citet{Dhoke&Paranjape}, or direct computation of a functional form for merger rates, relating to overdensity and the highest progenitor masses \citep{FakhouriMa}.

\subsection{Stochastic Star Formation}
\label{sec:stoch}

With short-timescale star formation events being difficult to model, it may be worthwhile implementing a stochastic description of the \ac{sfh}, which has been achieved on Myr timescales \citep{Tacchella2020}. This stochastic modelling is accomplished by computing the power spectrum of the \ac{sfh} through implementing variable gas inflows and molecular cloud formation while conserving the system's mass. This power spectrum captures the timescales of these processes, allowing one to constrain the lifetime of molecular clouds and star-forming regions. In our data, the power spectra of our predicted \ac{sfh} and \ac{zh} are weaker at high frequencies when compared with \ac{illtng} data. However, a network model in which we predict this spectrum directly may provide a statistical model for star formation on smaller timescales, potentially offering a stochastic amendment to observables.

We tested a network for central and for satellite galaxies, in which the same dark matter properties and network architecture were used to predict the amplitude of the Fourier transforms of the \ac{sfh} and \ac{zh}, i.e. the square root of their power spectra. The shape of the spectrum and the scaling of its amplitude with halo mass are well recovered, with similar levels of accuracy as the \ac{sfh} and \ac{zh} themselves. It is therefore possible to develop a model which could predict these power spectra and generate a stochastic signal to mitigate the errors in our predictions of \ac{sfh} and \ac{zh}. The correlations between residual Fourier transforms and luminosities and colours seen in \cref{sec:observables} suggest that this can help obtain more realistic galaxy SEDs.

\subsection{Metallicity}
\label{sec:metalpred}

We find that the complete networks roughly predict the mass-weighted stellar metallicity of all galaxies as a function of time, recovering the relations between stellar mass and metallicity, albeit with smaller scatter. High metallicity objects have their metallicity underestimated, which can be attributed to low sampling of the most metal-rich galaxies, or poor prediction of the star formation variability which weighs it.

The omission of time-dependent overdensity, circular velocity or radial skew undermines the ability to predict the scatter in the final metallicity and \ac{zh} variations. Overdensity and skew are environmental variables, providing measures of the concentration of the local mass distribution, and the density bias towards or away from the target subhalo, respectively. As functions of time, they measure the development of the most significant subhalo interaction events.

The influence of the skew on \ac{zh} scatter may be tracing the infall of subhalos, whose relation to the target's mass and environment adds to the likelihood of the halo progenitor hoarding star-forming or metal-rich gas. The importance of overdensity in determining metallicities would suggest that chemical enrichment owes to the abundance and properties of interacting subhalos. For centrals, such a relationship may be a consequence of strong correlations of overdensities with the metal richness of gas inflows and the intergalactic medium \citep{PengMaiolino}, while the tendency for overdensity to capture major interaction events which accelerate star formation \citep{LHuillier} may also contribute to the galaxy's metallicity in future. However, overdensity appears more important for satellite metallicities. This significance may be attributed to environmental quenching, which is not a common effect on central galaxies, whose quenching is strongly correlated with the galaxy's mass \citep{Bluck}. It is possible that the effect that overdensities have on central metallicities is captured by another shuffle group.

The skew, on the contrary, is weighted by the mass distribution and is independent of the scaling of the subhalo masses involved. Instead it can be considered a measure of the trajectory and frequency of subhalo interactions, most of which are minor. This could explain why overdensities are important to the galaxy's mass-weighted metallicity, but the skew has notable secondary effects on the metallicity history.

The circular velocity also influences the metallicity scatter, albeit, with lower significance. This quantity strongly correlates with subhalo mass, yet it has a strong connection to the timescale of internal collapse, and thus the onset of AGN feedback which expels star-forming gas \citep{JJDavies2020}, with higher values indicating rapid collapse and high halo concentration, thereby creating an environment which can alter the likelihood of metal production. This scaling with halo concentration, hence the star formation efficiency, is also dependent on environment and cosmic web distances \citep{Hellwing}, so this is far from an independent measure of these dynamics. Yet the omission of $\tilde{v}_\text{vir}$ distorts the predicted \ac{mhzr} scatter, suggesting the importance of dynamical measures of metal synthesis. Furthermore, shuffle group 3 contorts the distribution of metallicities on sample scales; a unique property of this shuffle group. While it is unclear whether this last distinction highlights a physical or a deceptive correlation, this could suggest that the dynamical nature of the galaxy-halo connection is the most independent of halo mass.

If the network is run with the baryon-biased maximum circular velocity in place of our virial approximation, the under-prediction of high metallicity objects is reduced, and the \ac{mzr} is fit more accurately. The full enrichment of these objects cannot be determined by our dark matter quantities and is additionally dependent on a purely baryonic phenomenon. Our model will undermine the metallicity scatter in a pure dark matter application. This result may prove more fruitful if our networks were trained on a semi-analytic model simulation, if this model returns suitable metallicities despite the dark matter determined rotation curve.

The dark matter half-mass radius, also a structural quantity, does not make a difference to predictions in metallicity, which has been shown to be the case at least at $z=0$ \citep{Lovell}. The physical size of the halo is evidently insignificant next to its central density. This quantity does, however, appear important for predicting gas mass and instantaneous star formation rate \citep{Lovell}, which in principle constrains galaxy growth in the long term, however this is already dominated by mass accretion and other variables.

\subsection{Stellar Mass \& Star Formation History}
\label{sec:masspred}

Building our networks with relatively few parameters, or randomising several quantities, does not adversely influence the predictions of the \ac{shmr} or star formation histories. The differences in scatter or halo-binned stellar mass distributions are minimal between runs with differently randomised training data.

Final halo mass and circular velocity are powerful parameters to determine the stellar mass. Historically, stellar mass and \ac{shmr} scatter depend greatly on the origin time and the halo mass accretion rate. The difference with satellites is that while the halo mass is most crucial to the stellar mass in the time-independent framework, it is the subhalo's mass accretion history which results in greater errors for the neural network as it is randomised. The properties and environment of the host can be attributed to the quantity of satellite galaxies and their evolution post-infall \citep{Bose}, therefore there may be a future study in which the assembly bias of the host halo is used to determine satellite formation history.

Quantities pertaining to the infall of satellite subhalos are also important for their \ac{sfh}. Though infall time is a significant factor in determining the \ac{sfh}, as this will restrict the onset of satellite physics, we argue that the relationship between $a_\text{max}$ and $a_\text{infall}$ is a valuable asset to our model; the two share the half-mass formation time as a denominator, and therefore this relates the time of infall to the accretion history. Strictly, if $a_\text{max} > a_\text{infall}$, then the subhalo continues to grow after becoming bound to a larger halo, advocating a greater star formation rate in the satellite phase. On the contrary, the infall time along with $t_\text{start}$ can constrain this relationship between central-phase accretion and post-infall star formation.

We have included starting time and scaled accretion times along the satellite's \ac{mpb}, yet the formation times affiliated with the main halo are known to constrain the population of satellite halos and the properties of the galaxies they host \citep{Artale, Bose}. While these times concerning the host halo are likely to be captured by its mass formation rate, evaluating this explicitly in our model may indicate the typical properties of satellite galaxies within an early or late forming halo.

Inspecting star formation histories, it appears that the time at which star formation begins to decline and the subsequent gradient of the \ac{sfh} are well matched for satellite galaxies. On the one hand, the mass ratio upon infall is likely to indicate the subsequent rate of stripping of the satellite galaxy, while the scaled infall time relates the time at which this ratio is achieved to the central-phase growth of the satellite galaxy. It would appear, however, that the deduction made by scaled time factors can be inferred by the combined growth histories of the subhalo and its host. The conclusions from \citet{Shi} indicating the clear distinction in the star formation rates linked to rapidly and slowly growing subhalos could be derived from their temporal implementation, while the merger histories which also correlate with their scaled formation time can be inferred from overdensity and skew histories.

It may be worthwhile in future designs to include properties relating directly to quenching, particularly given that this is unconstrained for central galaxies. A halo's internal velocity dispersion proves fruitful in classifying quenching central galaxies, being tightly related to black hole mass, while nearest-neighbour overdensity is important for quenching satellites \citep{Bluck}. As these are valid measures of the harassment of a star-forming region, as a time-dependent input they may correlate with dips or drops in \ac{sfh}. The difficulty is in obtaining a measure of velocity dispersion which is not affected by baryons.

\subsection{Modelling Observables}
\label{sec:obs}

Systematic differences between predicted \ac{sfh} and \ac{zh} may be mitigated in future by means of a stochastic measure of their fluctuations. We can train the same neural networks to predict the Fourier power spectra of these histories, correcting for the smaller amplitude of our predictions at high frequencies. We have shown that residual luminosities and line emission features correlate with this loss of amplitude, and therefore this correction could improve the accuracy of our spectra and photometry.

\subsection{Mock Surveys}
\label{sec:mocks}

Our model paves the way for large-volume high fidelity mocks and the construction of mock spectroscopic surveys with a consistent galaxy-halo connection. Such mocks can cover a large sample of galaxies and at a variety of redshifts. With respect to models that directly infer observational properties from dark matter properties, our model has the advantage that it links those observables with physical properties self-consistently. This expands the potential utility of these mocks in supporting the analysis of real surveys.

\section{Summary \& Conclusions}
\label{sec:conclusion}

We have developed a neural network model capable of synthesising the star formation rate and stellar metallicity of central and satellite galaxies as a function of cosmic time, using historical properties of the dark matter halo and environment as predictors. We have tested for the predicting power of certain dark matter quantities in determining aspects of the galaxy's evolutionary history, and computed observational properties from our predictions. The following statements summarise important aspects of our predictions:

\begin{itemize}
\item For each of our networks, designed separately for central and for satellite galaxies, the full diversity of star formation histories are accurately predicted. In \cref{sec:shmr,sec:sfh2} we see that geometries of the predicted \ac{sfh}s match those from the original \ac{illtng} merger trees, from those of continually growing star-forming galaxies to quenched satellites and high-mass galaxies. The integrated stellar mass from these historical growths are well correlated with their values in the \ac{illtng} data, and their mass from their merger trees. We show that our predicted star formation histories recover the complete \ac{shmr} in \cref{fig:shmr}.
\item This work includes the introduction of novel parameters, such as the skew of the radial mass distribution surrounding the target object, and recently devised quantities such as the scaled formation time from \citet{Shi}. These quantities have proven valuable in the prediction of our star formation and metallicity histories, and their relations with halo and galaxy masses have been outlined by the neural network.
\item The derived \ac{shmr} for either dataset is accurate and resilient to the removal of predicting features, i.e. it can be maintained by relatively few parameters. However, we show in \cref{sec:relimp2b} that omission of certain properties can cause subtle differences in the scatter of the \ac{shmr}, as indicated by multiple shuffled runs. Removing quantities related to subhalo growth rate and circular velocity have such an impact.
\item In spite of these fruitful outcomes, this network model seldom predicts star formation events which occur on short timescales, such as star formation bursts or rapid quenching. We see the absence of such features in \cref{fig:sfh} and the general lack of high frequency information. Consequently, numerical stellar mass estimates where these sharp changes are prominent are systematically smaller than their true value, and the absence of such features can impact the accuracy of observational properties, which is often seen in \cref{sec:observables}. This is a greater issue for central galaxies, where these features are more commonplace. We can, however, train an identical network to fit the power spectra of the \ac{sfh}s, shown in \cref{sec:stoch}, potentially offering stochastic corrections in future. Untested properties relating to the host of a satellite galaxy may prove valuable, however more general properties such as velocity dispersion are difficult to characterise independently of baryons.
\item In \Cref{sec:zh2}, we see that mass-weighted metallicity scatter is slightly underestimated, which in addition to the shortcomings of the star formation histories, may be attributed to the lack of baryonic influence on the rotation curve. We show in \cref{sec:relimp2b} that overdensity and skew, i.e. environmental measures, are paramount to the scatter in metallicity history; removing either of them affects their predictions significantly. Determining the variability in metallicity has been an issue in pre-existing studies \citep{Agarwal, Lovell}, however these include little detail around environmental history.
\item Despite some notable differences in metallicity, the predicted galaxy formation histories are observationally congruous to their hydrodynamical counterparts. While not predicting them directly, in \cref{sec:photometry} we show that the network recovers major observational statistics such as the mass-colour relations and colour bimodality of \ac{illtng}. H$\alpha$ line fluxes are recovered less accurately as these are sensitive to variations in recent star formation, yet they obey similar trends to the original data.
\item Mock observables are shown to be well matched to those emulated from \ac{illtng} data, despite inaccuracies such as smoother formation histories and lower metallicity scatter. Spectral energy distributions are computed more accurately for satellite galaxies than they are for central galaxies, as seen in \cref{sec:specsandlines}, likely owing to greater quenching. We show in \cref{sec:lumfourier} that larger residuals in high-frequency modes of the \ac{sfh} result in greater errors in luminosities, therefore a stochastic correction based on Fourier amplitudes may rectify this discrepancy. Other options include the introduction of dark matter quantities which may result in better predictions of formation history, or predicting star formation events on finer timescales.
\item In \cref{sec:relimp3}, while we motivate infall parameters as potentially important measures of satellite evolution, and properties of the cosmic web for central galaxies, we find that these have little isolated effect on our predictions, most likely owing to the inference of their effects from historical data. In high fidelity mocks, a higher volume of samples will be used to quantify the strength and concurrence of these correlations.
\end{itemize}

Ours is a practical model which can be used in high fidelity N-body simulations, statistically reproducing key relationships of the \ac{ghc}, and producing mock surveys of unprecedented size and complexity. A future publication will describe the predictions made using data from an N-body simulation, and the spectroscopic mocks which we aim to make publicly available.

\section*{Acknowledgements}

The authors would like to thank the anonymous referee for useful suggestions which led to significant improvements in the clarity of the manuscript. HGC wishes to thank the UKRI Science and Technology Facilities Council for funding this research. We wish to thank the IllustrisTNG project for access to their data and their JupyterLab workspace.

\section*{Data Availability}

The data supporting the findings of this research is currently available upon reasonable request from the main author. The data will later be made publicly available online.

\bibliographystyle{mnras}
\bibliography{bib}

\bsp	
\label{lastpage}
\end{document}